\documentclass[final,5p,times,twocolumn]{elsarticle}
\usepackage{amsmath,amssymb,amsfonts}
\usepackage{algorithmic}
\usepackage{graphicx}
\usepackage{textcomp}
\usepackage{hyperref}
\usepackage{color}
\usepackage{soul}
\newcommand{\hlc}[2][white]{{%
    \colorlet{foo}{#1}%
    \sethlcolor{foo}\hl{#2}}%
}
\usepackage{blindtext}
\usepackage{graphicx}
\usepackage[font=scriptsize,labelfont=bf,justification=justified]{caption}
\usepackage{hyperref}
\usepackage{amssymb}
\usepackage[ruled,linesnumbered]{algorithm2e}

\usepackage{arydshln}
\usepackage{verbatim} %for having comments
\usepackage{subfig} 
\usepackage{svg}
\usepackage{amsfonts}
\usepackage{amsopn}
\usepackage{amsthm}
\usepackage{xifthen}
\usepackage{pgfkeys}
\usepackage{multirow}
\usepackage{ragged2e}
\usepackage{xfrac}
\usepackage{pdfpages}
\usepackage{natbib}
\usepackage{cleveref}
\usepackage{float}
\usepackage[export]{adjustbox}
\usepackage{cleveref}
\usepackage{enumitem}
\usepackage{tabularx}

\usepackage{booktabs} 
\usepackage{ltablex}
\begin{document}
\begin{frontmatter}
\title{Appliance-level Flexible Scheduling for Socio-technical Smart Grid Optimization}
%%%%
%%%%
%%%%
\author{\corref{correspondingAuthor}Farzam Fanitabasi}
\address{Chair of Computational Social Science, ETH Zurich, Switzerland}
\cortext[correspondingAuthor]{Corresponding author. Address: Stampfenbachstrasse 48, 8092 Zurich, Switzerland. Email Address: \href{mailto:farzamf@ethz.ch}{\textit{farzamf@ethz.ch}}}
\author{Evangelos Pournaras}
\address{School of Computing, University of Leeds, Leeds, UK}
%%%%
%%%%
%%%%
\begin{abstract}
Participation in \hlc{residential} energy demand response programs requires an active role by the \hlc{consumers}.
\hlc{They} contribute flexibility in how they use their appliances as the means to adjust energy consumption, and \hlc{reduce demand peaks, possibly at the expense of their own comfort (e.g., thermal)}.
Understanding the collective potential of appliance-level flexibility for \hlc{reducing demand peaks} is challenging and complex.
\hlc{For instance, physical characteristics of appliances, usage preferences, and comfort requirements all influence consumer flexibility, adoption, and effectiveness of demand response programs.
To capture and study such socio-technical factors and trade-offs, this paper contributes a novel appliance-level flexible scheduling framework based on consumers' self-determined flexibility and comfort requirements.
By utilizing this framework, this paper studies (i) consumers usage preferences across various appliances, as well as their voluntary contribution of flexibility and willingness to sacrifice comfort for improving grid stability, (ii) impact of individual appliances on the collective goal of reducing demand peaks, and (iii) the effect of variable levels of flexibility, cooperation, and participation on the outcome of coordinated appliance scheduling.
Experimental evaluation using a novel dataset collected via a smartphone app shows that higher consumer flexibility can significantly reduce demand peaks, with the oven having the highest system-wide potential for this.
Overall, the cooperative approach allows for higher peak-shaving compared to non-cooperative schemes that focus entirely on the efficiency of individual appliances}.
The findings of this study can be used to design more cost-effective and granular \hlc{(appliance-level)} demand response programs in participatory and decentralized Smart Grids.
\end{abstract}
%%%%
%%%%
%%%%
\end{frontmatter}
%%%%
%%%%
%%%%
\section{Introduction}
\label{S:introduction}
The European 2030 climate and energy framework has set three key targets for the year 2030:
At least 40\% reduction in greenhouse gas emissions, 27\% share of renewable energy, and 27\% improvement in energy efficiency from the 1990 levels~\cite{EU2030}.
\hlc{Meanwhile, the global electricity consumption in residential sector, which accounts for $30-40\%$ of the total energy usage, is ever increasing}~\cite{alberini2011response,torriti2014review}.
\hlc{This forces the utility companies to expand their energy generation and transmission capacity to address the occasional peak demands.}~\cite{li2019climate}.
\hlc{Energy demand response programs aim to match the demand to the available supply to reduce/prevent peak energy demands, thus, improving grid stability, avoiding blackouts, and reducing pollution}~\cite{agnetis2013load,kohlhepp2019large,yahia2020multi}.
\hlc{In the residential sector, this matching can be realized by consumers (households) \textit{adjusting} and altering their electricity usage from their normal consumption patterns, either by changing the appliance time-of-use (load-shifting), or by reducing the energy consumption.
This adjustment represents the level of consumer flexibility, and can be in response to price changes over time, or incentive rewards designed to induce lower electricity use at times of high demand, or when system reliability is jeopardized}~\cite{qdr2006benefits,palensky2011demand,adika2014autonomous}. \\

\hlc{Given the prominent role of residential consumers in participatory demand response programs, it is crucial to understand their energy usage habits, preferences, comfort requirements, and flexibility.
Recent research also highlight the socio-technical aspects of Smart Grids, and the importance of designing residential demand response programs in a more bottom-up, and consumer-centric manner}~\cite{mckenna2018simulating,mammoli2019behavior}.
\hlc{Where consumer's age, income, household size, and working hours, as well as increasing attention to privacy, self-determination, and autonomy, influence the adoption and effectiveness of demand response programs}~\cite{yilmaz2019sensitive}.
\hlc{Previous work on residential demand response programs commonly formulate consumers' flexibility and preferences as a set of constraints}~\cite{baldauf2015smart,yao2016real}.
\hlc{However, consumers' flexibility and preferences are not always hard constraints}~\cite{yilmaz2019sensitive}, rather they vary based on the appliance type, time-of-use, individual characteristics and behavior~\cite{gyamfi2013residential}, conventions~\cite{powells2014peak}, monetary incentives~\cite{verbong2013smart}, and social practices involving the appliances~\cite{torriti2017understanding}.
\hlc{Hence, various studies consider consumer flexibility and preferences as objective functions, leveraging multi-objective optimization of appliance-level schedules, aiming to minimize consumer inconvenience, electricity cost, and peak load, simultaneously}~\cite{yahia2020multi,wang2015robust}.
\hlc{Furthermore, depending on their intrinsic interests and incentives, consumers can decide to further cooperate with the demand response program (i.e., voluntarily contribute flexibility), and sacrifice their comfort for the collective goal of reducing demand peaks.
To the best of authors' knowledge, this trade-off between consumer cooperation and demand peak reduction has not been previously studied.
Lastly, most literature on residential appliance-level load scheduling analyses single households, thus neglecting the collective potential of coordinated appliance scheduling among multiple households, as well as the impact of individual appliances, and their usage features on potential load-shifting and peak-shaving}~\cite{jordehi2019optimisation,yahia2020multi}. \\

\hlc{Designing an appliance-level flexible scheduling framework that effectively captures such socio-technical trade-offs and coordinate energy demand across multiple households in a decentralized manner is the challenge that this paper tackles.}
To this end, this paper proposes a novel appliance-level energy scheduling framework that relies on consumers' self-determined flexibility and comfort requirements, to regulate energy demand.
\hlc{By utilizing the framework, this paper studies (i) consumers usage preferences across various appliances, as well as their voluntary contribution of flexibility and willingness to sacrifice comfort for improving grid stability, (ii) impact of individual appliances on the collective goal of reducing demand peaks, and (iii) the effect of variable levels of flexibility, cooperation, and participation on the outcome of coordinated appliance scheduling.}
\hlc{In the proposed framework, the collective goal is to prevent demand peaks (peak-shaving/clipping) by means of load-shifting.}
whilst for individual consumers, the objective is to maximize comfort by using their appliances at the desired time~\cite{chavali2014distributed}.
Preventing demand peaks can be achieved by leveraging consumer flexibility in appliance usage, and uniformly distributing the demand across the day \hlc{(i.e., minimizing demand variance)}~\cite{palensky2011demand,spiliotis2016demand,hassan2013impact}.
In this setting, consumer flexibility is considered to be the contribution of alternative appliance usage schedules.
For instance, multiple schedules as a result of shifting an appliance usage earlier or later in time from the intended usage time~\cite{spiliotis2016demand,zhai2019appliance,d2015demand,mohsenian2010autonomous}.
This flexibility creates a degree of freedom for coordination within the framework to optimize the selection between these alternatives in a way that reduces the peak-load~\cite{powells2014peak,mckenna2018simulating,gyamfi2013residential,verbong2013smart}. \\

\hlc{However, coordinating consumers' schedules for reducing demand peaks remains challenging due to several factors.}
The two objectives of maximizing comfort and reducing demand peaks can be opposing, as certain appliance usages might be delayed (or advanced), thus lowering consumers' comfort~\cite{pilgerstorfer2017self,pournaras2014decentralized}.
Moreover, \hlc{given consumer flexibility and usage preferences}, such coordination requires selecting a subset of discrete schedules based on a quadratic cost function (minimizing demand variance), which is an NP-hard combinatorial optimization problem~\cite{pilgerstorfer2017self,de2018complexity}. 
This calls for approximation mechanisms to find a near-optimal and computationally feasible solution~\cite{molzahn2017survey,petersen2013taxonomy}.
This paper addresses these challenges by introducing a decentralized network of autonomous scheduling agents, each representing a \hlc{consumer (i.e., a household with multiple appliances).}
These agents cooperatively coordinate to select a subset of consumers' schedules to reduce demand peaks.
To optimize agents' selections, this paper applies the I-EPOS (\emph{Iterative Economic Planning and Optimized Selections})~\cite{pilgerstorfer2017self} system, to perform fully decentralized, privacy-preserving, and multi-objective combinatorial optimization.
Depending on the appliance and its automation level, these coordinated schedules can be executed via smart appliances~\cite{fanti2019cooperative}, \hlc{home energy management system (HEMS)}~\cite{sattarpour2018multi}, \hlc{or presented as recommendation to the consumers}~\cite{haider2016review}. \\

In summary, the contributions of this paper are the following: 
(i) A novel appliance-level scheduling framework based on consumers' self-determined flexibility and comfort requirements, \hlc{performing multi-objective optimization of appliance schedules across multiple households, aiming to reduce demand peaks.}
(ii) A data-driven analysis of appliance-level socio-technical factors, such as cooperation level, and unfairness that influence consumers' flexibility and \hlc{the collective demand peak reduction.}
(iii) \hlc{Impact analysis of individual appliances on reducing demand peaks (with oven being the most impactful), and} a quantitative comparison to related work which reveals that in comparison to improving appliance efficiency, flexible coordinated scheduling can further \hlc{reduce demand peaks in Smart Grids}.
(iv) A new dataset on flexible scheduling of appliances by residential consumers. 
The rest of this paper is outlined as follows: 
Sections~\ref{S:relatedWork} and~\ref{S:overview} summarize related work, and provide an overview of the framework operations, respectively.
Section~\ref{S:applianceScheduling} introduces the flexible scheduling model, and Section~\ref{S:IEPOS} illustrates the distributed combinatorial optimization system.
In Section~\ref{S:expMethod} the experimental methodology of the paper along with the dataset, survey, and mobile application are illustrated.
Section~\ref{S:evaluation} shows the experimental evaluation.
Finally, Section~\ref{S:conclusion} concludes this paper and outlines future work.
%%%%
%%%%
%%%%
	 \begin{figure*}[!htb]
	 \centering
	 \includegraphics[width=0.9\textwidth]{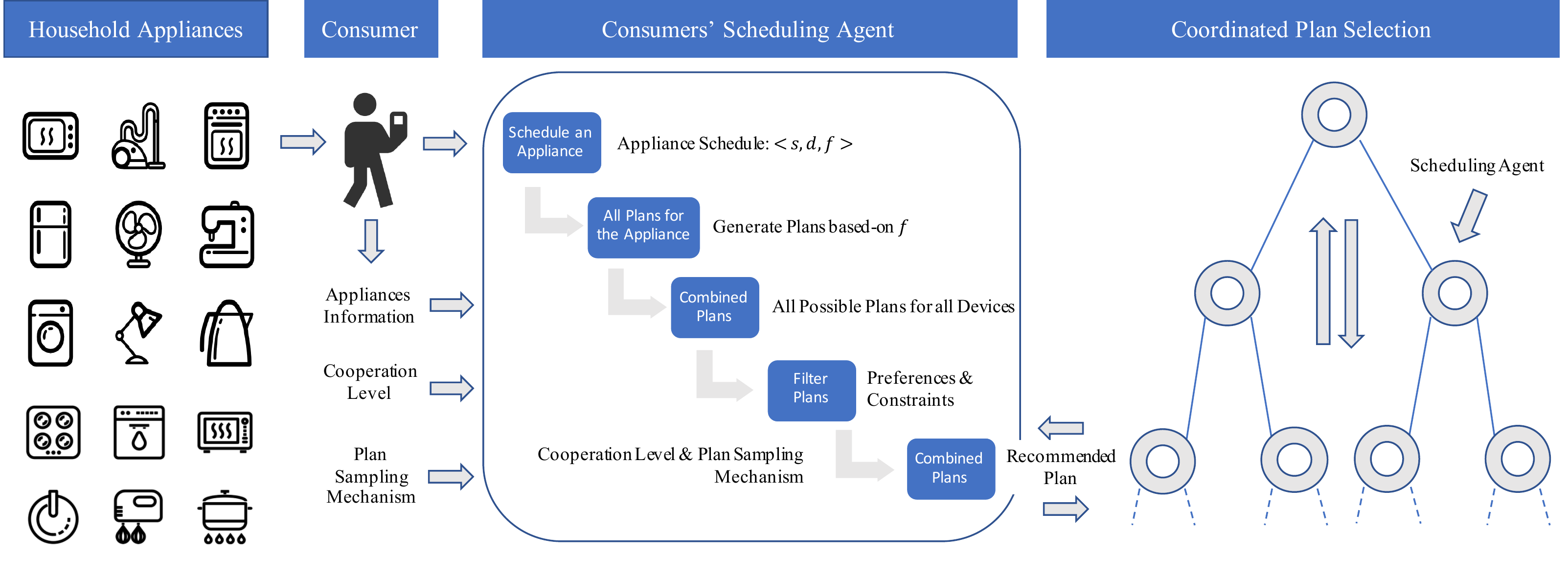}
	 \caption{Overall scenario: 
	 The consumer $u$ initializes its scheduling agent and adds appliances information, usage constraints, and \hlc{cooperation level.}
	 Then, the consumer submits the appliance schedules with their flexibility values. 
	 The scheduling agent generates all possible plans for each appliance based on Algorithm~\ref{A:planGeneration}, combines the plans for all appliances, and applies consumers' constraints and preferences on the plans.
	Then, the scheduling agent samples plans based on \hlc{a local} sampling mechanism, and provides them as input to I-EPOS. 
	\hlc{Based on consumers' cooperation level,} I-EPOS coordinates and selects a subset of these plans with the aim of \hlc{reducing demand peaks}. 
	Finally, the selected plan is presented to the consumer via the scheduling agent \hlc{to be executed by smart appliances, home energy management systems (HEMS), or used as recommendations.}}
	\label{fig:Scenario}
	\end{figure*}
%%%%
%%%%
%%%%
\section{Related Work}
\label{S:relatedWork}
Demand response programs for Smart Grids have been subject to extensive research~\cite{jordehi2019optimisation,haider2016review}.
Several studies attempt to model markets and pricing schemes to coordinate consumers' \hlc{energy demand with the supply.}
Examples include game-theoretic approaches~\cite{mohsenian2010autonomous,rahimi2010overview,tushar2015three}, heuristic evolutionary algorithms~\cite{logenthiran2012demand}, and agent-based techniques~\cite{jordan2018better,pournaras2017self}. 
\hlc{Previous studies identify} residential appliance scheduling via demand response programs as a \hlc{viable} approach for improving Smart Grid efficiency and utilization~\cite{kohlhepp2019large,haider2016review,d2015demand,adika2014autonomous}.
Often, such programs adopt \hlc{load-shifting to achieve peak-shaving, or valley-filling on the aggregated demand}~\cite{torriti2014review,kohlhepp2019large}. 
To perform load-shifting, the consumers' flexibility in appliance usage is calculated, often using the following two approaches:
(i) Estimation of flexibility based on extrapolated consumption data~\cite{d2015demand,agnetis2013load,spiliotis2016demand,zhai2019appliance,hassan2013impact,ji2017demand,drysdale2015flexible,kwon2014assessment,dyson2014using,joe2012optimized,macdougall2016applying}.
(ii) Simulating the operating times of appliances, as well as usage habits~\cite{murray2016understanding,adika2014autonomous,d2015demand,yin2016quantifying,olivieri2014evaluation,ali2014demand,bartusch2014further,saele2011demand,nistor2015capability, halvorsen2001flexibility,petersen2013taxonomy}. \\

\hlc{Utilizing consumer flexibility, optimal load scheduling of household appliances based on multi-objective optimization have been subject of previous studies.
Such studies leverage mixed integer programming}~\cite{molzahn2017survey,muhsen2019domestic,yahia2018optimal}. \hlc{heuristic methods}~\cite{molzahn2017survey,de2018complexity,soares2014multi}, \hlc{or non-exact approximation methods}~\cite{molzahn2017survey,setlhaolo2014optimal,shakouri2017multi,setlhaolo2015optimal,setlhaolo2016combined}.
\hlc{Among which, Yahia et al.}~\cite{yahia2018optimal} \hlc{utilizes various multi-objective optimization schemes to simultaneously minimize electricity cost and consumer inconvenience, to reduce demand peak, across multiple households.
Setlhaolo et al.}~\cite{setlhaolo2014optimal} \hlc{proposes a non-linear model to co-optimize consumption cost and carbon emissions, though they do not address reducing demand peaks.
Weighted sum multi-objective optimization models have also beed studied to minimize electricity cost and consumer inconvenience based on time-of-use pricing}~\cite{setlhaolo2016combined}, \hlc{or incentives}~\cite{shakouri2017multi}.
\hlc{However, these studies often only consider a single household, thus ignoring the collective potential of coordinated appliance scheduling among multiple households, or do not consider consumers' self-determined flexibility contribution and cooperation in the collective goal of reducing demand peaks}~\cite{molzahn2017survey,yahia2018optimal}.
\hlc{Additionally, within the setting of coordinated appliance scheduling, the impact of individual appliances on the collective goal of reducing demand peaks requires further studies}~\cite{jordehi2019optimisation,yahia2020multi}.
To address this, the framework introduced in this paper leverages consumers' self-determined flexibility, comfort requirements, \hlc{and cooperation across multiple households} to study the collective potential of appliance load scheduling for reducing peak demands. \\

\hlc{Additionally, given usage constraints and number of discrete options/schedules, coordinating appliance schedules across multiple households for reducing demand peaks is non linear (minimizing demand variance), NP-Hard, and combinatorially complex}~\cite{de2018complexity}.
Distributed optimization, and genetic algorithms are utilized to cope with this complexity, and to approximate a near-optimal solution between consumers' demand and the available energy supply of the utility company~\cite{molzahn2017survey}. 
\hlc{Genetic algorithms formulated as multi-objective optimization problems are utilized to simultaneously minimize: electricity cost and consumer dissatisfaction}~\cite{soares2014multi}, \hlc{cost and delays in appliance usage}~\cite{muralitharan2016multiobjective}, \hlc{and cost and consumer inconvenience}~\cite{muhsen2019domestic}.
However, these studies do not address the distributed and privacy sensitive nature of appliance usage data.
To this end, this paper utilizes and expands I-EPOS to provide a distributed, privacy-preserving coordination and optimization scheme for consumers' schedules~\cite{pilgerstorfer2017self}. \\

Lastly, recent research also highlight the socio-technical aspects of Smart Grids and emphasized the need to design demand response programs in a more bottom-up, and consumer-centric manner~\cite{mckenna2018simulating,mammoli2019behavior}.
\hlc{These studies argue that} socio-technical aspect such as age, income, household size, and working hours, as well as increasing attention to privacy, self-determination, and autonomy influence the \hlc{adoption and effectiveness} of demand response programs~\cite{yilmaz2019sensitive}.
Additionally, consumer flexibility does not only depend on the appliances and monetary incentives~\cite{verbong2013smart}, but on individual characteristics \hlc{(i.e., intrinsic motivations such as environmental awareness)}~\cite{gyamfi2013residential}, conventions~\cite{powells2014peak}, and social practices involving the appliances~\cite{torriti2017understanding}.
However, flexibility estimations based on extrapolated or simulated data often fail to capture such socio-technical factors, and do not account for the scenarios where consumers' behavior deviates from the norm (e.g., having guest, or going on holidays).
Thus, this paper utilizes a personalized scheduling agent for each consumer.
Using this scheduling agent, consumers directly determine their data and privacy preference, schedules, flexibility, usage preference, scheduling constraints, and \hlc{voluntary contribution of flexibility for reducing demand peaks}, on a daily basis\footnote{While this approach does require a higher level of engagement from consumers, recent research has shown that values such as increased control and autonomy can improve the acceptance and adoption of such programs\cite{verbong2013smart,ellabban2016smart,murtagh2014qualitative}.}.
%%%%
%%%%
%%%%
\begin{table}[t]
\centering
\caption{Mathematical notations used in this paper} 
\scriptsize
\begin{tabular}{ll} 
\toprule
\textbf{Notation} & \textbf{Meaning} \\
\midrule
$U$ & Set of all consumers \\
\addlinespace
$u \in U$ & Consumer $u$ \\
\addlinespace
$H_u$ & Set of all appliances of consumer $u$ \\
\addlinespace
$h_i \in H_u$ & Appliance $i$, belonging to consumer $u$ \\
\addlinespace
$\langle s,d,f \rangle$ & Representation of an appliance schedule \\
\addlinespace
$\langle v_1,...,v_{n} \rangle$ & Representation of an appliance plan \\
\addlinespace
$P_u$ & Set of all plans of consumer $u$ \\
\addlinespace
$p_{u,i} \in P_u$ & Plan $i$ of consumer $u$ \\
\addlinespace
\hlc{$k$} & \hlc{Number of plans submitted to I-EPOS for each consumer} \\
\addlinespace
$f_L \colon p_{u,i} \rightarrow \mathbb{R}$ & \hlc{Local cost function (discomfort)}  \\
\addlinespace
$f_G \colon \mathbb{R}^{n} \rightarrow \mathbb{R}$ & \hlc{Global cost function (minimizing demand variance)} \\
\addlinespace
$a_1^t$ & Aggregate plans of agents at root at iteration $t$ \\
\addlinespace
$a_u^{t}$ & Aggregate plans of agents beneath $u$ at iteration $t$  \\
\addlinespace
$p_{u,o} \in S_u$ & Original plan of agent $u$ with no flexibility \\
\addlinespace
$p_{u,s}^t \in S_u$ & Selected plan for agent $u$ at iteration $t$ \\
\addlinespace
$p_{u,s}^F\in S_u$ & Final selected plan for agent $u$ by I-EPOS \\
\addlinespace
$\lambda := 0\leq \lambda \leq1$ & \hlc{Consumer's cooperation level} \\
\addlinespace
$D^t$ & Average discomfort of agents at iteration $t$ \\
\addlinespace
$\Psi^t$ & Unfairness across agents at iteration $t$ \\
\bottomrule
\end{tabular}
\label{t:Maths}
\end{table}
%%%%
%%%% 
%%%%
\section{Overview \& Framework Operations}
\label{S:overview}
Figure~\ref{fig:Scenario} \hlc{illustrates the framework, and its four main operations.}
(i) \hlc{Each consumer interacts with a prototyped digital assistant running as an app on a smartphone, acting as a personal scheduling agent.}
Via this scheduling agent, consumers indicate their appliances' schedules \hlc{(i.e., starting time and duration of operation)}, and usage constraints, such as which appliances should not be used in parallel\footnote{Scheduling is a well-established approach in literature\cite{fanti2019cooperative,chavali2014distributed,fathi2013adaptive,rokni2018optimum,wang2018green,lahon2019contribution,de2005multi,maliah2017collaborative,georgeff1988communication}, and has been utilized in several real-world application domains\cite{pilgerstorfer2017self,pournaras2017self,barbati2012applications}.}.
Consumers also indicate the discomfort level they are willing to tolerate as an indicator of flexibility, such as to what extent \hlc{the appliance usage} can shift in time from the desired operational time.
This discomfort level determines how willing they are to contribute to the collective goal, i.e. \hlc{reducing demand peaks.}
(ii) The consumers submit their appliance schedules and flexibility \hlc{for the next day} to the scheduling agent\footnote{Previous studies have shown that daily energy usage follows a semi-repetitive manner\cite{halvorsen2001flexibility}.
Hence, the consumers can choose to schedule once for each day, and only modify them if required.
This greatly reduces the scheduling effort on part of the consumers.}.
By leveraging flexibility, multiple alternative energy consumption patterns are generated using the scheduling agent (Section~\ref{S:applianceScheduling}), each called a \textit{plan}\footnote{Note that a schedule specifies the intended appliance usage. and is defined independent of the appliance energy consumption. 
But the plans specify the exact energy consumption of the appliance during its use.}.
\hlc{The scheduling agent then removes the conflicting plans based on consumers' constraints (e.g., not showering while the oven is on).}
(iii) \hlc{Reducing demand peaks, computationally modeled as the minimization of a non-linear cost function, i.e. variance of total energy demand across the day}~\cite{sattarpour2018multi,pournaras2017self}, \hlc{requires coordination among consumers' schedules, which is an NP-hard combinatorial optimization problem}~\cite{pilgerstorfer2017self,de2018complexity}.
This calls for approximation mechanisms to find a near-optimal solution at low computational cost~\cite{molzahn2017survey,petersen2013taxonomy}. 
\hlc{This framework relines on} a decentralized network of autonomous scheduling agents, each representing \hlc{a residential household with multiple appliances.}
\hlc{These agents interact and} coordinate to select a subset of consumers' schedules to reduce demand peaks\footnote{Crucially, the selected plans should not violate the consumers' comfort expectations, or the physical constraints of the Smart Grid, for instance, the power generator limits~\cite{spiliotis2016demand}.}..
To optimize agents' selections, this paper utilizes and extends the I-EPOS (\textit{Iterative Economic Planning and Optimized Selections})~\cite{pilgerstorfer2017self} system, to perform fully decentralized, privacy-preserving, and cooperative combinatorial optimization.
(iv) Finally, the coordinated plans are submitted back to the scheduling agent.
Note that this paper studies a range of personal appliances, ranging from highly automated (washing machine), to low automation (TV).
Depending on the appliance and its automation level\footnote{Such as allowing for direct control, and interruptible/deferrable operation.}, the recommended usage plans can be executed automatically via the home energy management system (HEMS) (in case of washing machines), or \hlc{presented} directly to the consumer as recommendations (in case of TV).
The above framework can be operationalized in different scales depending on the available computational resources.
Ranging from cooperative micro grids (e.g., smart building) where individual appliances form atomic units, to hierarchical scenarios where the lowest levels are households that aggregate to form neighborhoods, and districts.
In the next sections the above scenario is illustrated in more detail.
%%%%
%%%%
%%%%
\section{Flexible Appliance Scheduling}
\label{S:applianceScheduling}
The utilized mathematical notations are presented in Table~\ref{t:Maths}. 
A consumer $u \in U$ schedules the usage of appliance $h\in H_u$. 
A schedule is defined as: $\langle s,d,f \rangle$, where $s$ is the preferred starting time, $d$ is the duration in minutes, and $f$ is the appliance flexibility in minutes. 
This flexibility means that the consumer is willing to shift its original starting time $s$, either earlier or later, by $f$ minutes. 
A plan $i$ of consumer $u$ ($p_{u,i}~\in P_u$) is defined as a sequence of real values $\langle v_1,...,v_{n} \rangle$ of size $n=1440$ (i.e., 24*60, number of minutes in a day). 
Each $v_j$ represents the energy consumption of the appliance on the $j^{th}$ minute of the day.
For each schedule, using the flexibility provided by the consumer, the scheduling agent generates all possible plans ($P_u$) based on Algorithm~\ref{A:planGeneration}. 
A schedule with flexibility $f$ results in multiple plans, where the earliest plan starts at $s-f$, and the latest plan at $s+f$. \\

Associated with each plan is its \textit{discomfort}, \hlc{caused by possible changes from the intended} starting time of the appliance usage.
Intuitively, the closer a plan is to the preferred starting time, the lower the discomfort it imposes on the consumer, which in turn increases its \hlc{acceptance and adoption.}
\hlc{This discomfort is calculated by a local cost function}: $f_L(p_{u,i})$. 
The plan derived by $f=0$ is represented as $p_{u,o}$ and its discomfort is 0 (i.e., $f_L(p_{u,o})=0$).
An example of such a process is shown in Table~\ref{SampleAS}.
%%%%
%%%%
%%%%
\SetInd{0.5em}{0.5em}
\begin{algorithm}[t]
\footnotesize
\DontPrintSemicolon
\SetAlgoLined
\KwIn{schedule $S:\langle s,d,f \rangle$ for appliance $h$, time granularity $g$ (default: minutes), scheduling horizon (default: 24h)}
\KwOut{$P$: list of all plans for schedule $S$}
/* number of plans based on flexibility\;
$k \leftarrow 2*f~/~g + 1$ \;
/* list of all plans including the plan with $f=0$\;
Initialise array $P$ of size $k$\;
/* calculate the plan length\;
$n \leftarrow \text{24h}/g$  (based on the scheduling horizon) \;
$e \leftarrow$ energy consumption of $h$ per $g$\;
\For{$i$ in $0$ to $k-1$}{
	/* create a vector of 0's size n for each plan\;
	$V_i:~<v_1,...,v_n>$ $\leftarrow 0$\;
	\For{$j$ in $0$ to $d-1$}{
	/*$s-f$ should be $\geq 0$ as day-before schedules are not allowed
	 $v_{s-f+j} \leftarrow e$\;
		}
	$f \leftarrow f-1$\;
	add $V_i$ to $P$\;
	}
\caption{Plan generation}
\label{A:planGeneration}
\end{algorithm}
%%%%
%%%%
%%%%
	\begin{table}[!htb]
	\centering
	\caption{Consumer $u$ schedules the kettle usage at 6pm, for 10' and has 2' flexibility: $\langle \text{18:00},10',2' \rangle$. 
	This means that the preferred starting time is 18:00 but the consumer is flexible for the start time to be between 17:58 to 18:02. 
	Hence, 5 different plans are generated, the first one starting from 17:58 and ending at 18:08 and the last one starting from 18:02 and ending at 18:12. 
	Assume that $f_L(p_{u,i})$ calculates the discomfort as the absolute difference of the plan start time from the preferred starting time. 
	The 5 \hlc{possible starting times and their discomfort} are listed below:}
	\scriptsize
	\begin{tabular}{lll} 
	\toprule
	\textbf{Appliance} &\textbf{Plan Start Time} & \textbf{Discomfort} \\
	\midrule
	Kettle & 17:58 & 2 \\
	Kettle & 17:59 & 1 \\
	Kettle & 18:00 & 0 \\
	Kettle & 18:01 & 1 \\
	Kettle & 18:02 & 2 \\ 
	\bottomrule
	\end{tabular}
	\label{SampleAS}
	\end{table}
%%%%
%%%%
%%%%
\subsection{Constraints \& Preferences}
\label{S:consPrefPlanSpace}
\hlc{The plans that do not adhere to consumers constraints are removed from $P_u$.}
\hlc{For instance, if the consumer} does not wish to shower while the oven is on, or use the washing machine after 9 pm. 
Moreover, consumers' preferences for different plans are measured by \hlc{their discomfort}, with higher preference given to plans with lower discomfort.
If the consumer schedules multiple appliances, the scheduling agent combines the plans from appliances. 
This process is performed as follows: 
assume consumer $u$ schedules $h_i$ with $f=p$, and $h_j$ with $f=q$.
Each schedule generates $2f+1$ plans\footnote{$f$ number of plans with starting time before $s$, another $f$ with starting time after $f$, plus the plan with $f=0$.}, hence, $4pq+2p+2q+1$ combined plans.
\hlc{Each combined plan represents the aggregated energy consumption of the two appliance plans.
The discomfort of a combined plan is the average discomfort of the plans.}
This process is performed by the scheduling agent. 
%%%%
%%%%
%%%%
\subsection{Consumers' Cooperation Level}
The collective goal of the proposed framework \hlc{(reducing demand peaks throughout the day)} is measured by the global cost function: $f_G(C)$, where $C = \{p_{u,s}~|~\forall u \in U)\}$ is the set of all selected plans by the consumer, and $p_{u,s} \in P_u$ is the selected plan that consumer $u$ intends to execute. 
This can be achieved by minimizing the variance of consumers' total energy demand across the day~\cite{sattarpour2018multi,pournaras2017self}.
To do so, the framework leverages consumers' flexibility to shift their appliance usage in time.
Such an approach requires collective action and cooperation by the consumers, and can reduce consumers' comfort, due to the shift in appliance usage time.
Thus, the consumers can determine their \textit{\hlc{cooperation level}} in two phases: (i) plan generation (scheduling), and (ii) plan selection.
%%%%
%%%%
%%%%
\subsubsection{Plan Generation (Scheduling) Phase}
\label{S:PGP}
For a given schedule $\langle s,d,f \rangle$, if consumer $u$ determines a high flexibility value, then the scheduling agent creates a high number of possible plans $|P_u| >> 1$. 
A high number of possible plans increases the computational complexity and the storage capacity requirements of the agent\footnote{For example, in a scenario with 100 agents, each with 10 plans, the solution space size is $10^{100}$.}.
Assume that consumer $u$ needs to submit $k < |P_u|$ plans to the demand response program, and the plans in $P_u$ are sorted based on their \hlc{discomfort.}
\hlc{The consumer} can utilize various plan sampling mechanisms on $P_u$ in order to indicate \hlc{the cooperation level}.
A simplified version of the social value orientation theory~\cite{mcclintock1989social} is used to study the range of non-cooperative \hlc{(minimizing discomfort) to fully cooperative (minimizing demand peaks) behaviors.} 
A non-cooperative consumer samples $k$ plans with the lowest discomfort, while the fully cooperative consumer samples $k$ plans with \hlc{highest discomfort.}
Intuitively, due to the popularity of certain actions, the cooperative consumers provide the demand response program with more diverse plans. 
For instance, using the oven is a very common between 6-7pm~\cite{torriti2017understanding}.
Thus, a cooperative consumer that \hlc{shifts the oven usage} before 6pm or after 7pm, \hlc{is more likely to contribute significantly} to reducing demand peaks.
%%%%
%%%%
%%%%
\subsubsection{Plan Selection Phase}
\hlc{Based on the sampled plans, $\forall u \in U$, plan $p_{u,s}$ is selected (Section}~\ref{S:IEPOS}) \hlc{according to the following equation:}
%%%%
%%%% 
%%%%
\begin{equation}
\label{eq:autonomousPresPlanning}
\arg\min_{s=1}^{k} \Big( (1-\lambda)*f_G(C) + \lambda*f_L(p_{u,s}) \Big)
\end{equation}
%%%%
%%%% 
%%%%
in which $\lambda := 0 \leq \lambda \leq 1$ is the weight assigned to the \hlc{discomfort} and is determined by each consumer, indicating the cooperation level in the plan selection phase, and $k$ is the number of plans for consumer $u$
A higher $\lambda$ value indicates \hlc{a non-cooperative consumer (minimizing discomfort).}
Various incentivisation schemes (e.g., monetary rewards) can be used to motivate consumers to set lower $\lambda$ values.
%%%%
%%%% 
%%%%
\section{Coordinated Plan Selection}
\label{S:IEPOS}
\hlc{Reduction of demand peaks by minimizing the variance of consumers' total energy demand across the day, requires coordination between different consumers and their energy usage plans}~\cite{jordehi2019optimisation,haider2016review,pilgerstorfer2017self}.
\hlc{To perform this coordination and optimize the selection of plans,} consumers' scheduling agents employ the I-EPOS system~\cite{pilgerstorfer2017self} as a \hlc{multi-agent}, fully decentralized, self-organizing, and privacy-preserving combinatorial optimization mechanism\footnote{Accessible online at \href{https://github.com/epournaras/EPOS}{https://github.com/epournaras/EPOS} (last accessed: April 2020)} .
\hlc{Each I-EPOS agent has a set of discrete energy consumption plans, generated by the corresponding scheduling agent.}
I-EPOS coordinates and selects a subset of consumers' plans aiming to minimize the variance of the total energy demand across the day \hlc{aiming to reduce/prevent demand peaks ($f_G$: MIN-VAR)}~\cite{pilgerstorfer2017self}.
\hlc{To this end,} I-EPOS agents self-organize in a tree-topology~\cite{pournaras2010self} as a way of structuring their interactions, \hlc{facilitate the cost-effective aggregation of plans, as well as performing coordinated optimization and decision-making.}
I-EPOS performs consecutive learning iterations. 
Each iteration has two phases: the bottom-up (leaves to root) phase and top-down (root to leaves) phase. 
During the bottom-up phase of each iteration, agent $u$ selects the plan $p_{u,s}$ which satisfies the following \hlc{multi-objective optimization objective:}
%%%%
%%%%
%%%%
	\begin{equation}
	\begin{aligned}
	\label{eq:IEPOSExpanded}
	p_{u,s}^t = \arg\min_{s=1}^{k}~\Bigg((1-\lambda)\Big({\sigma^2}\big(a_1^{t-1} - a_u^{t-1} + a_u^t \\
	-~p_{u,s}^{t-1} + p_{u,s}^t\big)\Big) + \lambda\Big(f_L(p_{u,s}^t)\Big)\Bigg)
	\end{aligned}
	\end{equation}
%%%%
%%%%
%%%%
where $\sigma^2$ is the variance function, and $t$ is an iteration of I-EPOS. 
$a_1^{t-1} = \sum_{u=1}^{|U|} p_{u,s}^{t-1}$ is the \textit{aggregate plan} at iteration $t-1$ of the selected plans of all agents, summed up at the root.  
$a_u^{t-1}$ and $a_u^t$ shows the aggregate plan calculated by summing up the selected plans of agents in the branch below agent $u$, at iterations $t-1$ and $t$, respectively.
$p_{u,s}^{t-1}$ and $p_{u,s}^t$ are the selected plans of agent $u$ at iteration $t-1$ and $t$, respectively. 
Finally, consumers' \hlc{cooperation level} is included in the objective via the $\lambda$ parameter. \\
\hlc{The coordination among consumers is performed by each agent aggregating its selected plan $p_{u,s}^t$ with $a_u^t$, and communicating the aggregated plan to the parent agent in the tree topology.}
In the top-down phase, agents are notified about $a_1^{t-1}$.
After the final iteration $F$ is completed, $p_{u,s}^F$ is presented to the consumer by the scheduling agent of consumer $u$.
Moreover, at each iteration $t$, the \hlc{average discomfort} $D^t$ across all agents, is calculated as: 
	\begin{equation}
	\label{AVGDiscomfort}
	D^t  = \frac{1}{|U|} \sum_{u}^U (f_L(p_{u,s}^t)). 
	\end{equation}
\hlc{Additionally,} at each iteration $t$, the deviation of discomfort values across all agents is referred to as unfairness $\Psi^t$, calculated as:
	\begin{equation}
	\label{eq:unfairness}
	\Psi^t = \sqrt{ \frac{1}{|U|} \sum_{u}^U (f_L(p_{u,s}^t))^2 - (\frac{1}{|U|} \sum_{u}^U f_L(p_{u,s}^t))^2 }.
	\end{equation}
Further elaboration on I-EPOS is out of the scope of this paper and the interested reader is referred to previous work~\cite{pilgerstorfer2017self}. 
%%%%
%%%%
%%%%
\subsection{Complexity, Optimality, and Privacy}
The computational and communication complexity of the above algorithm are $O(kt\log |U|)$ and $O(t\log |U|)$, respectively. 
Where $k$ is the maximum number of plans per agent, $t$ is the number of iterations, and $|U|$ is the number of agents/consumers~\cite{pilgerstorfer2017self}.
This allows for efficiency and scalability to scenarios with higher number of agents.
\hlc{In regards to the addressed objective function if this paper (NP-Hard combinatorially complex optimization), previous research}~\cite{nikolic2019structural} \hlc{has revealed that compared to the state-of-art, I-EPOS has a superior performance profile.
Regarding optimality, in a solution spaces of size $2^{20}$ I-EPOS empirically finds solutions in the top 33\% of all solutions in the first learning iteration, and top 3.35\% in the last learning iteration}~\cite{pilgerstorfer2017self}.
Lastly, the preservation of privacy is based on the fact that each agent only shares the aggregated plan to the parent agent in the tree topology, and does not disclose preferences, $p_{u,s}$, $P_u$, or $\lambda$.
%%%%
%%%%
%%%%
\section{Experimental Methodology}
\label{S:expMethod}
\hlc{This section introduces the data collection, plan sampling mechanisms, and the experimental design of this paper.}
%%%%
%%%% 
%%%% 
\subsection{Data Collection}
\label{S:dataset}
%%%%
%%%%
%%%%
\begin{figure}[!htb]
	 \includegraphics[width=0.45\textwidth]{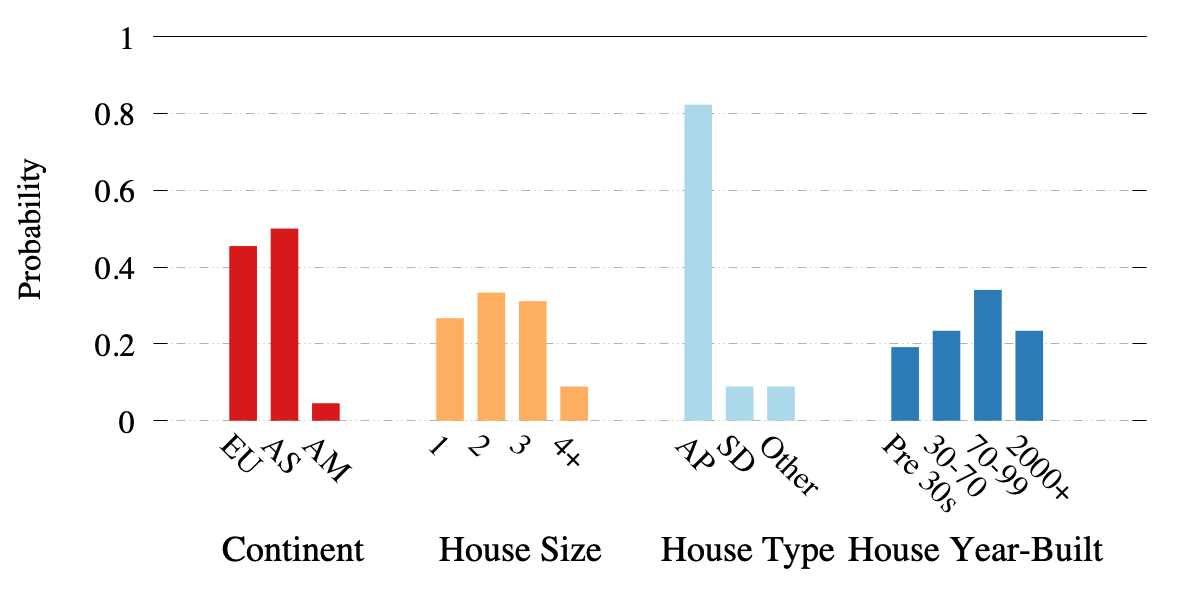}
	 \caption{Participant information in the collected dataset. 
	 Continent refers to where participants currently reside in. 
	 EU stands for Europe, AS for Asia, and AM for the Americas. 
	 House size represents the number of bedrooms. 
	 House type shows the distribution of participants living in apartments (AP), semi-detached housed (SD) or other types of houses. 
	 House year-built shows the distribution of the year the participants' house was built. 
	 These features are specifically chosen to provide a comparison baseline with the REFIT dataset\cite{murray2017electrical}, and to assist in determining consumption profiles}
	\label{Demography}
	\end{figure}
%%%%
%%%%
%%%%
\hlc{Modeling and evaluating appliance-level flexible scheduling is challenging, as such low granularity and privacy-sensitive data are usually not (openly) available by power utilities.}
Moreover, large-scale social studies \hlc{linked with consumers actual energy consumption behavior} at the appliance level are usually costly, over-regulated, and require complex interventions by power utilities. 
To overcome such limitations that have significantly restricted the scope of earlier research~\cite{zhai2019appliance,d2015demand,pournaras2014decentralized}, two datasets are combined to make this research feasible: 
(i) A new real-world dataset\footnote{Available online:~\cite{Fanitabasi2019Figshare} (last accessed: April 2020)} \hlc{collected in a field study using an Android mobile application, based on the illustrated scenario} in Section~\ref{S:overview}.
This dataset contains appliance usage schedules and flexibility from 51 participants\footnote{The participants were recruited through a cross-university campaign.} across four days, from 4 to 8 February 2018, \hlc{together with complementary survey data.}
This dataset contains appliance usage schedules and flexibility from 51 participants across four days, from 4 to 8 February 2018, \hlc{together with a complementary survey.}
To the best of the authors' knowledge, this dataset is the first pilot study which addresses consumers' self-determined flexibility, as well as \hlc{the cooperation level at the appliance level.}
Table~\ref{planDist} illustrates the distribution of the schedules and plans across different household appliances. 
This dataset is further analyzed in Section~\ref{S:evaluation}.
(ii) The state-of-the-art REFIT dataset~\cite{murray2017electrical}, \hlc{including electrical load measurements from 20 households at aggregate and appliance level, timestamped and sampled at 8-second intervals.}
%%%%
%%%%
%%%%
\begin{table}[!htb]
\centering
\caption{Distribution of schedules and plans among appliances}
\scriptsize
\begin{tabular}{llll}
\toprule
\textbf{Appliance} & \textbf{\# of Schedules} & \textbf{\# of Plans} & \textbf{Percentage (\%)}\\
\midrule
Computer & 62 & 6237& 14.76\% \\
\addlinespace
Dish Washer & 40 & 4320&  9.52\% \\
\addlinespace
Hob & 44 & 6516&  10.47\% \\
\addlinespace
Kettle & 80 & 5896&  19.04\% \\
\addlinespace
Oven & 97 & 10026&  23.09\% \\
\addlinespace
Tumble Dryer & 15 & 2184&  3.57\%  \\
\addlinespace
Washing Machine & 82 & 11570&  19.52\% \\
\addlinespace
Total & 420 & 46749&  100\% \\
\bottomrule
\end{tabular}
\label{planDist}
\end{table}
%%%%
%%%%
%%%%
\subsubsection{Mobile Application}
\label{S:plannerApp}
\begin{figure}[!htb]
	\centering
	 \includegraphics[width=\linewidth]{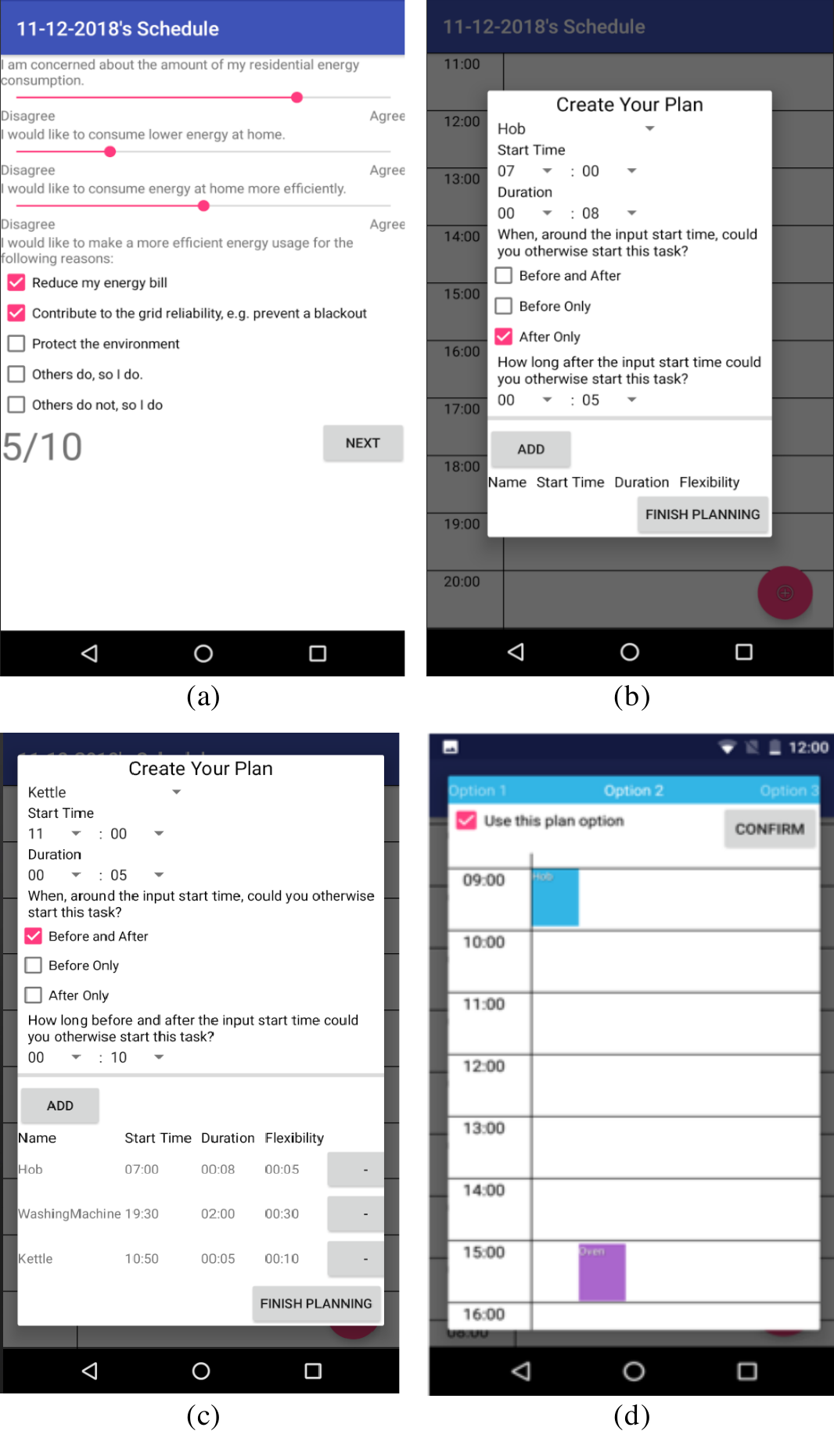}
	 \caption{Developed Android mobile application as the consumers' scheduling agent: (a) Sample of survey questions, (b) The schedule submission phase. (c) Schedules for different appliances. (d) Selected plan after I-EPOS execution.}
	\label{plannerAppScreens}
\end{figure}

An Android mobile application\footnote{Accessible online at \href{https://github.com/epournaras/EPOS-Smart-Grid-Scheduler}{https://github.com/epournaras/EPOS-Smart-Grid-Scheduler} (last accessed: April 2020)} (Figure~\ref{plannerAppScreens}) was developed and distributed among the participants, \hlc{as their personal scheduling agent.}
\hlc{The scheduling agent} is in charge of receiving the schedules, generating the plans, enforcing constraints and preferences, interacting with I-EPOS, \hlc{and presenting the selected plans to the consumers.}
%%%%
%%%% 
%%%%
\subsubsection{Participants Survey}
\label{survey}
\hlc{Accompanying the data collection, a survey was conducted to gain better insight into participants' energy consumption behavior, flexibility, and cooperation.}
The participants were invited to answer questions about demographic, household, and energy usage preferences, \hlc{such as discomfort threshold, willingness to schedule appliances, and their cooperation level\footnote{The survey is conducted according to the GDPR guidelines, and the identity of participants remains anonymous.}.} 
The detailed questionnaire and responses are presented in Appendix C.
%%%%
%%%% 
%%%%
\subsubsection{Appliance Energy Consumption}
\label{S:consProf}
Figure~\ref{Demography} shows some general information about the participants and their living situation.
\hlc{Each participant is considered a consumers, and based on the} household information, the scheduling agent matches each consumer \hlc{in the collected dataset} to the closest household in the REFIT dataset~\cite{murray2017electrical}, \hlc{to estimate consumers' appliance energy consumption.}
To estimate the consumption profile of a given consumer, the scheduling agent utilizes a linear scoring model of four household features: number of occupants, the built year, the house type, and the number of bedrooms.
These features have been empirically assigned weights of 0.533, 0.267, 0.133, and 0.067, respectively.
The assignment of these weights is based on the importance of each feature on household energy usage~\cite{baker2008improving,kim2018characteristics}
\footnote{However, as this paper mostly studies the consumers' scheduling and socio-technical factors affecting it, these weights do not affect the findings.
The linear combination is used for the sake of simplicity and interpretability.}.
Appendix A illustrates the household matching process and the distribution of profiles among consumers in more detail.
%%%%
%%%% 
%%%%
%%%%
%%%% 
%%%%
\subsection{Plan Sampling Mechanism}
\label{S:sampTech}
This paper utilizes five different plan sampling mechanisms, each sampling 10 plans from $P_u$ for all consumers.
Each sampling mechanism indicates a particular \hlc{cooperation level for consumers} at the plan generation (scheduling) phase (Section~\ref{S:PGP}).
\begin{enumerate}[leftmargin=*]
\item \textbf{Top Ranked}: Non-cooperative \hlc{\textit{(selfish)}} consumer; 
samples the top 10 plans from the plan space ($P$) with the lowest \hlc{discomfort}. 
\item \textbf{Top Poisson}: Semi-non-cooperative \textit{(semi-selfish)} consumer; 
samples 10 plans from the plan space, skewed towards lower \hlc{discomfort}. 
Modeled via a Poisson distribution\footnote{\label{note1}With a $\lambda$ parameter of 2 for the Poisson distribution.} on $P_u$ ordered by increasing \hlc{discomfort}.
\item \textbf{Uniform}: Balanced \textit{(fair)} consumer; 
samples 10 plans uniformly distributed across the plan space.
\item \textbf{Bottom Poisson}: Semi-cooperative \textit{(semi-altruistic)} consumer; 
samples 10 plans skewed towards the higher \hlc{discomfort} from the plan space. 
Modelled via a Poisson distribution\textsuperscript{\ref{note1}} on $P$ ordered by decreasing \hlc{discomfort}.
\item \textbf{Bottom Ranked}: \hlc{Fully-}cooperative \textit{(altruistic)} consumer; 
samples the top 10 plans from the plan space with the highest \hlc{discomfort}.
\end{enumerate}
%%%%
%%%%
%%%%
\begin{figure*}[!htb]
\centering
\subfloat[Appliance usage throughout the day with no flexibility]{\includegraphics[width = 0.45\textwidth]{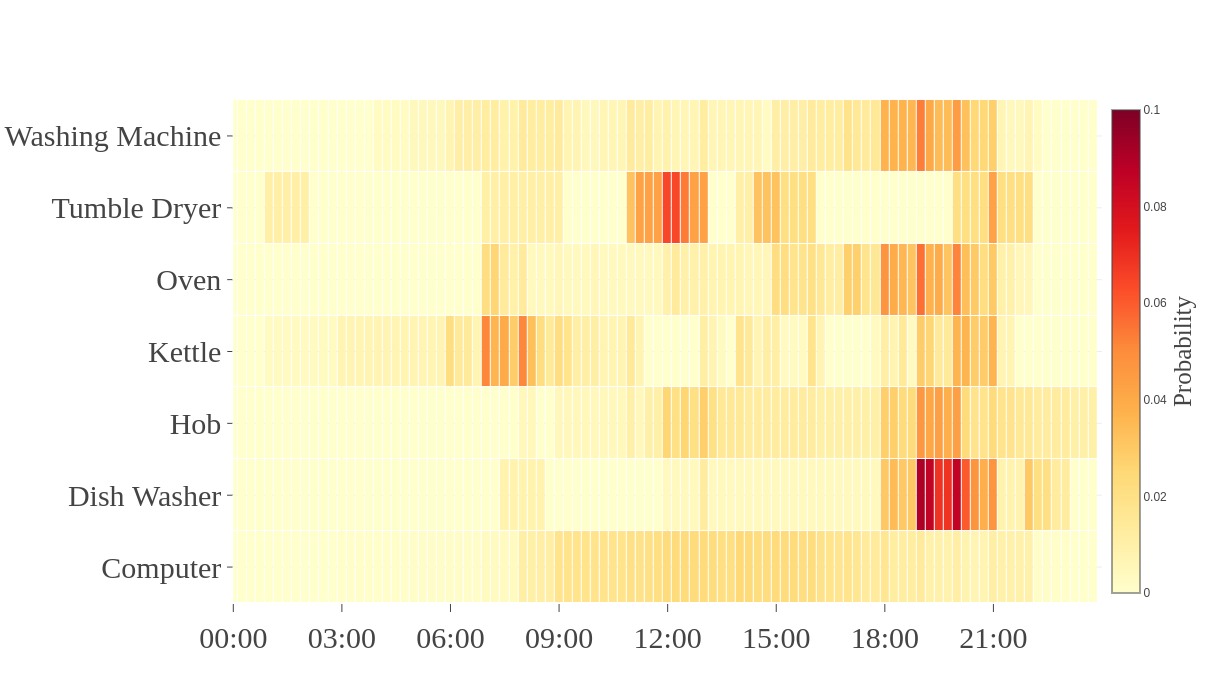}}
\subfloat[Appliance usage throughout the day given consumer flexibility]{\includegraphics[width = 0.45\textwidth]{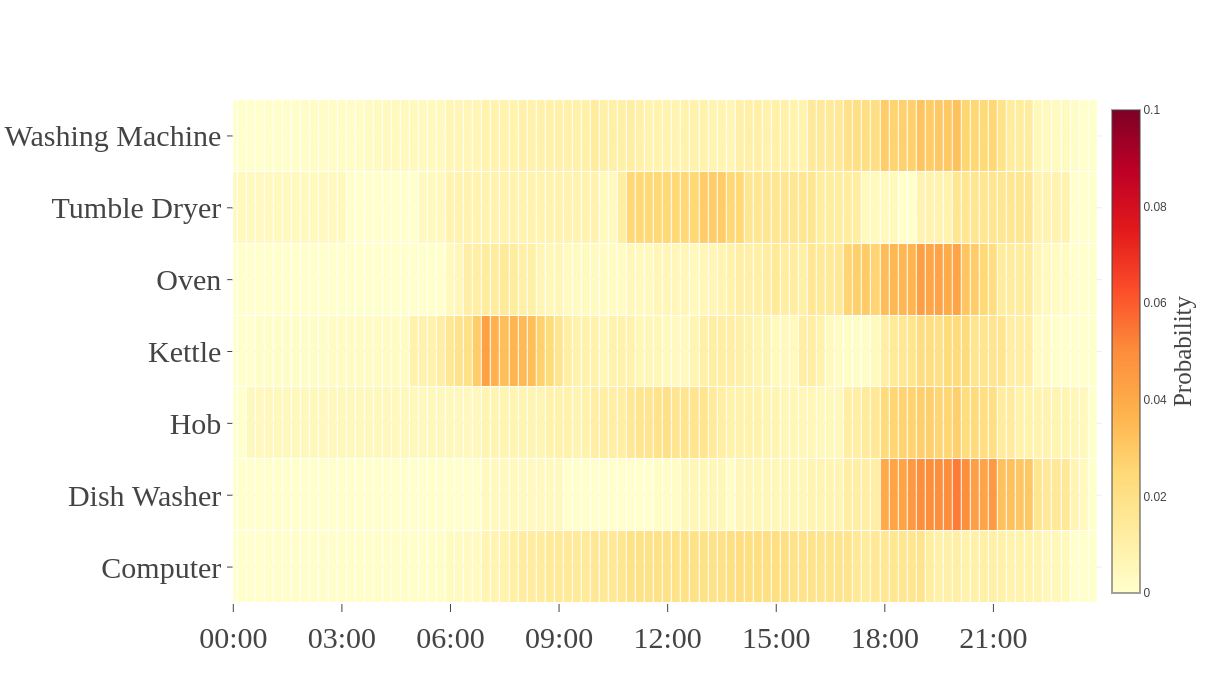}}
\caption{Comparison of appliance usage probability throughout the day with and without consumer flexibility}
\label{fig:Flex96}
\end{figure*}
%%%%
%%%% 
%%%%
\begin{figure*}[!htb]
\centering
\subfloat[Density of usage duration for various appliances across all schedules.]{\includegraphics[width = 0.93\textwidth]{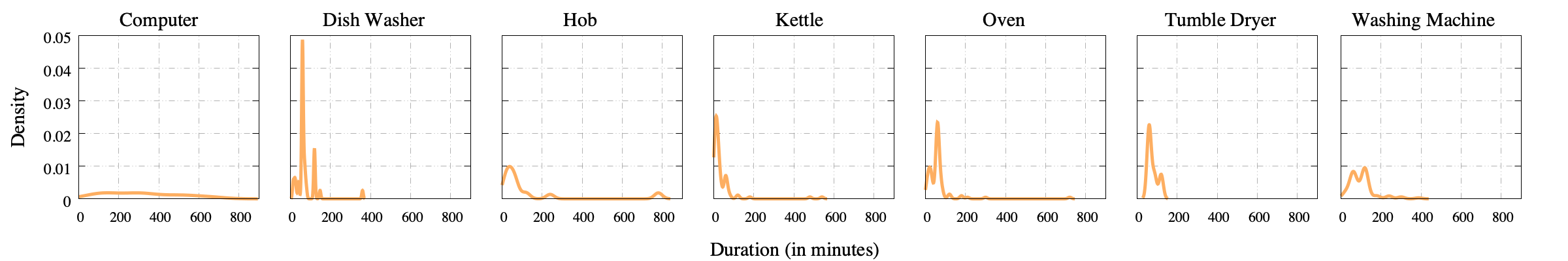}} \\ 
\subfloat[Density of usage flexibility for various appliances across all schedules.]{\includegraphics[width = 0.93\textwidth]{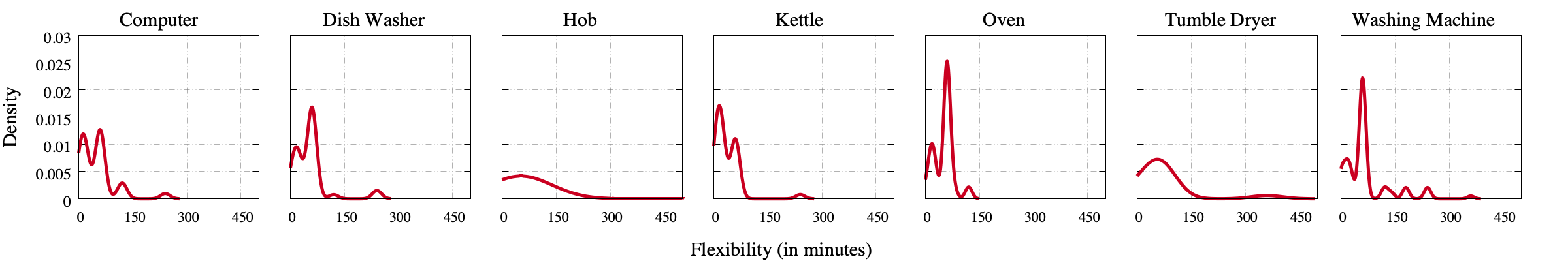}} \\
\subfloat[Density of relative flexibility for various appliances across all schedules.]{\includegraphics[width = 0.93\textwidth]{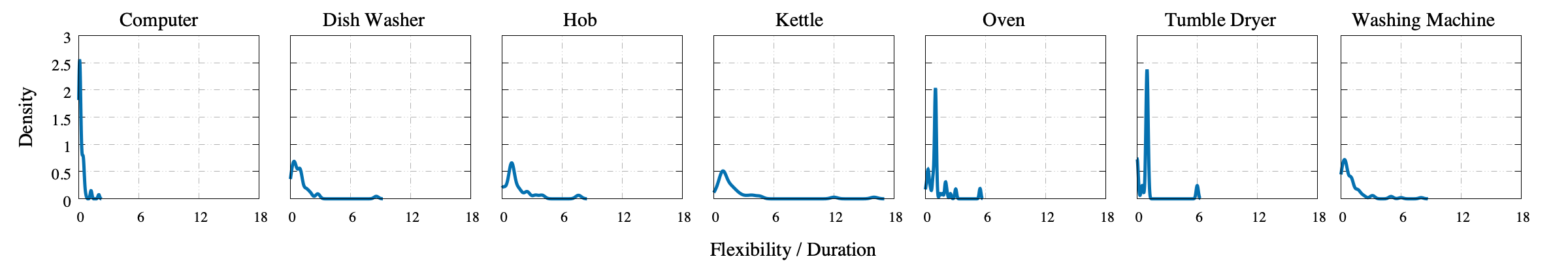}} \\
\caption{Density of duration, flexibility, and relative flexibility of various appliances.}
\label{fig:applianceKDE}
\end{figure*}
%%%%
%%%%
%%%%
\subsection{Experimental Design}
\label{S:Methodology} 
Table~\ref{Setting} illustrates the I-EPOS parameters used for the experiments.
\hlc{Each experiment is executed 10 times, and the reported results are averaged across all simulations.}
In each simulation of I-EPOS, the agents are randomly assigned to a position in the tree topology. 
The topology is a balanced binary tree. 
The consumers schedules were collected during 4 days. The first 3 days with 51 and the last day with 50 consumers\footnote{On the last day one of the consumers did not schedule.}.
The presented results are the average across the four days. 
The $\lambda$ parameter \hlc{(cooperation level)} is determined based on the survey results of each consumer, specifically the question P7 in Appendix C: 
``\textit{I would like to accept discomfort to make more efficient energy usage."}
The plan sampling mechanisms (Section~\ref{S:sampTech}) are used as a system-wide setting for all scheduling agents.
The scheduling agent samples and provides 10 plans to I-EPOS ($k=10$)\footnote{In Appendix B more experiments with 5 and 100 plans are shown to illustrate the effect of $k$ on the demand response program.}.
The plan dimensions (24*60) are the number of minutes in a day, where the value on each dimension shows the total energy usage on that minute by the consumer. 
The \hlc{discomfort is normalized and calculated as the distance between the $p_{u,o}$ and $p_{u,s}$, with the plan derived from $f=0$ having discomfort 0, the furthest plan ($s \pm f$) from the original starting time has discomfort 1.} 
The global cost function is MIN-VAR, which minimizes the variance of consumers' total energy demand, hence reducing demand peaks.
Below is the list of experiments and their methodology.
%%%%
%%%%
%%%%
\begin{table}[t]
\centering
\caption{I-EPOS parameters used in experiments.}
\scriptsize
\begin{tabular}{ll} 
\toprule
\textbf{Parameters} & \textbf{Value} \\
\midrule
Number of Executions & 10 \\
\addlinespace
Number of Iterations & 50 \\
\addlinespace
Number of Agents & 50-51 \\
\addlinespace
Number of Plans per Agent & 10 \\
\addlinespace
$\lambda$ \hlc{(Cooperation Level)} & Consumer Determined \\
\addlinespace
Plan Dimensions & 24*60 (1440)\\
\addlinespace
Network Topology & Balance Binary Tree \\
\addlinespace
Local Cost Function & Discomfort (shift in start time) \\
\addlinespace
Global Cost Function & MIN-VAR (minimizing variance) \\
\bottomrule
\end{tabular}
\label{Setting}
\end{table} 
%%%%
%%%% 
%%%%
\subsubsection{Flexible Appliance Scheduling \& Reducing Demand Peaks}
These experiments study the effect of consumers' flexibility \hlc{and cooperation} in scheduling their appliances \hlc{from three perspectives: (i) reducing demand peaks and demand variance, (ii) average discomfort, and (iii) unfairness.} 
Each agent sets its own $\lambda$ value \hlc{(cooperation level)} provided by the corresponding consumer (Section~\ref{S:Methodology}).
In addition, three other fixed system-wide values of $\lambda =0.0$, $0.5$, and $1.0$ are evaluated.
The experiments are repeated across the five different plan sampling mechanisms. 
%%%%
%%%%
%%%%
\subsubsection{Impact of Individual Appliances on Reducing Demand Peaks}
\hlc{These experiments} study the impact of individual appliances on the \hlc{collective goal of reducing demand peaks.
Due to consumers' preferences, appliance type (TV or computer usage), or costly hardware, the scheduling agent cannot control all the appliances.
Thus, understanding the load-shifting potential of various appliances for reducing demand peaks is essential for providing effective and efficient demand response solutions.}
To calculate this impact, 7 experiments are performed, each time excluding one of the appliances from the demand response program \hlc{by setting the flexibility of the excluded appliance to 0.
Furthermore, these experiments analyze existing appliance-usage features to determine whether such features can be utilized (individually or in combination) to calculate the importance, and the impact of various appliances on the demand response program.
For each appliance, the studied features include: average flexibility, average duration per use, average usage duration per day, relative flexibility, energy consumption, and percentage of plans in the dataset.}
%%%%
%%%%
%%%%
\subsubsection{Increased Efficiency vs. Flexible Coordinated Scheduling}
\label{S:M:IEOFS}
Earlier research has studied the impact of more efficient use of appliances on the Smart Grid;
specifically the increased energy efficiency of kettles if consumers avoid overfilling\cite{murray2016understanding}.
\hlc{To compare the reduction of demand peaks between the two approaches of ``increased energy efficiency" and ``flexible coordinated scheduling"}, the following methodology is used:
Scenario (a) ``Efficient Kettle": the \hlc{potential energy savings of the kettle from a previous study}~\cite{murray2016understanding} are applied to the consumers of the collected dataset, during peak hours (6:30-8:30 and 19:30-21:30). 
Scenario (b) ``Flexible Kettle": the I-EPOS experiments are performed by setting the flexibility of all appliances to 0, except the kettle. 
\hlc{Using this methodological approach}, in both scenarios, the two systems can only use the kettle to \hlc{reduce demand peaks.}
%%%%
%%%%
%%%%
\subsubsection{Varying Adoption of Recommended Plans}
Consumer participation is necessary for demand response programs to achieve their targets\cite{hassan2016framework}.
Given consumer autonomy over the \hlc{execution of recommended plans by the framework}, their participation and adoption level greatly affect how well the \hlc{framework achieves its targets}~\cite{hassan2016framework}.
This effect is analyzed by utilizing the following methodology:
The consumers are sorted in descending order, based on their $\lambda$ value. 
The reduced adoption is calculated by changing the $\lambda$ value of the top $n$-percent of the consumers with $\lambda \neq 1$ to 1. 
The consumers with $\lambda=1$ purely \hlc{minimize their discomfort, hence, they do not provide any alternative plans and effectively do not participate in the demand response program.}
Utilizing the above methodology, this experiment studies the necessary level of participation and adoption \hlc{by consumers to still achieve effective demand peaks reduction.}
%%%%
%%%% 
%%%%
\section{Results and Findings}
\label{S:evaluation}
This section illustrates the results and findings \hlc{based on the experimental designs in Section}~\ref{S:expMethod}.
%%%%
%%%% 
%%%%
\subsection{Dataset Analysis}
\label{S:datasetAnal} 
Figures~\ref{fig:Flex96}a and~\ref{fig:Flex96}b \hlc{show the probability of appliance usage across the days without and with consumer flexibility. 
As illustrated,} the inclusion of flexibility makes the likelihood of appliance usage more spread-out across the day.
For instance, the \hlc{usage probability the dish washer without flexibility is very high around 7pm} (Figure~\ref{fig:Flex96}a).
However, this probability is more distributed across 6-8pm when flexibility is included.
Figure~\ref{fig:applianceKDE} shows the density distribution\footnote{Calculated using the Gaussian kernel density estimator with the nrd0\cite{scott2015multivariate} algorithm for determining the bandwidth.} \hlc{of duration, flexibility, and relative flexibility ($flexibility/duration$) for all schedules.
For instance}, while the computer has the longest average duration (300') out of all appliances, its flexibility is relatively low (47'). 
On the other hand, the oven has a low duration on average (52'), however, it has a relatively high flexibility (60'). 
\hlc{Thus,} the oven can contribute more to distributing the energy demand across the day than the computer.
%%%%
%%%% 
%%%%
\begin{table}[t]
\centering
\caption{Relative flexibility of appliances throughout the day (Median): 
The morning hours are between 00:00 - 08:59, mid day between 09:00 - 16:59, and evening between 17:00 - 23:59. 
This splitting is made based on the common demand patters throughout the day~\cite{richardson2010domestic}}
\scriptsize
\begin{tabular}{lllll} 
\toprule
\textbf{Appliance} & \textbf{Morning} & \textbf{Mid Day} & \textbf{Evening} & \textbf{Average} \\
\midrule
Computer & 0.125 & 0.2 & 0.175 & 0.16\\
\addlinespace
Dish Washer & 0.46 & 1.64 & 1 & 1.03\\
\addlinespace
Hob & 4 & 1.66 & 1 & 2.22\\
\addlinespace
Kettle & 1.61 & 0.95 & 1.81 & 1.45\\
\addlinespace
Oven & 2.14 & 0.85 & 1 & 1.33\\
\addlinespace
Tumble Dryer & 0.66 & 0.7 & 1 & 0.78\\
\addlinespace
Washing Machine & 0.25 & 0.62 & 0.68 & 0.51\\
\bottomrule
\end{tabular}
\label{T:flexDay}
\end{table} 
%%%%
%%%% 
%%%%
Table~\ref{T:flexDay} illustrates the changes in the relative flexibility of the appliances throughout the day.
Additionally, Table~\ref{T:scheduleDay} in Appendix B illustrates the distribution of appliance schedules throughout the day.
%%%%
%%%% 
%%%%
The values of $\lambda$, \hlc{as an indicator of consumers' cooperation level,} are based on their answer to question P7 (Appendix C).
The participants are assigned one of the \hlc{5 discrete values between 0 to 1, with the resulting distribution:  $\lambda=$1 (5.88\%), 0.75 (35.3\%), 0.5 (27.5\%), 0.25 (21.6\%), 0 (27.5\%).}
\hlc{Intuitively, a $\lambda$ value of $1$ means that the consumer exclusively minimizes discomfort, while a $\lambda$ value of $0$ means the consumer is only concerned reducing demand peaks.
The average $\lambda$ value specified by the participants is $0.48$.}
%%%%
%%%% 
%%%%
\begin{figure}[!htb]
\centering
\subfloat[\hlc{Variance of consumers' total demand across the day}]{\includegraphics[width = 0.45\textwidth]{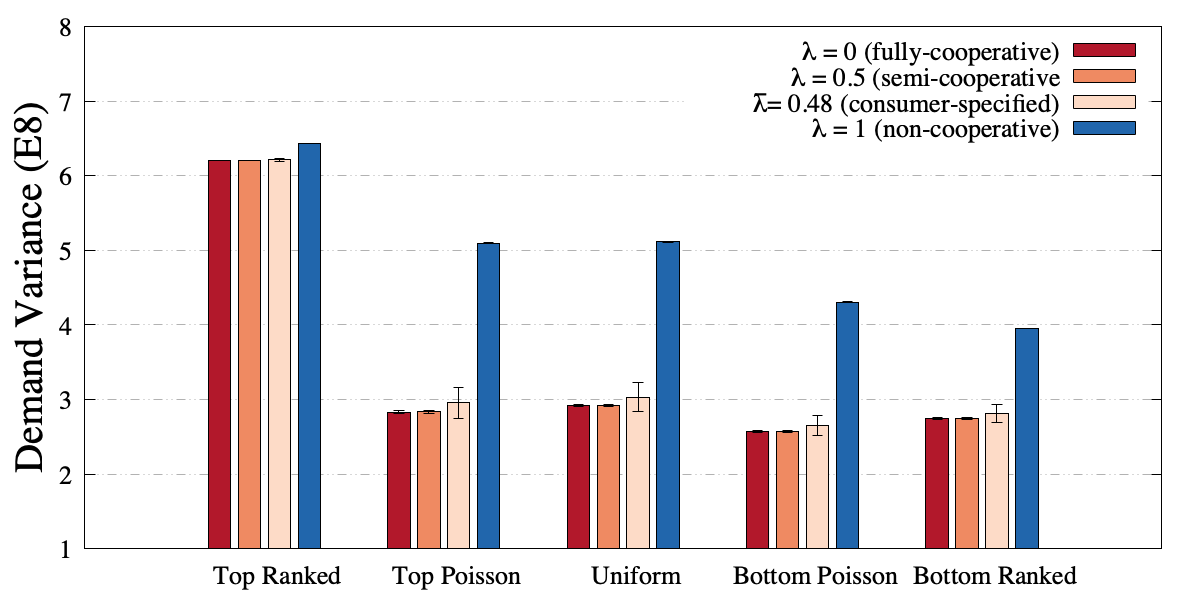}} \\ 
\subfloat[\hlc{Average discomfort} (Equation~\ref{AVGDiscomfort}) \hlc{of consumers' selected plans}]{\includegraphics[width = 0.45\textwidth]{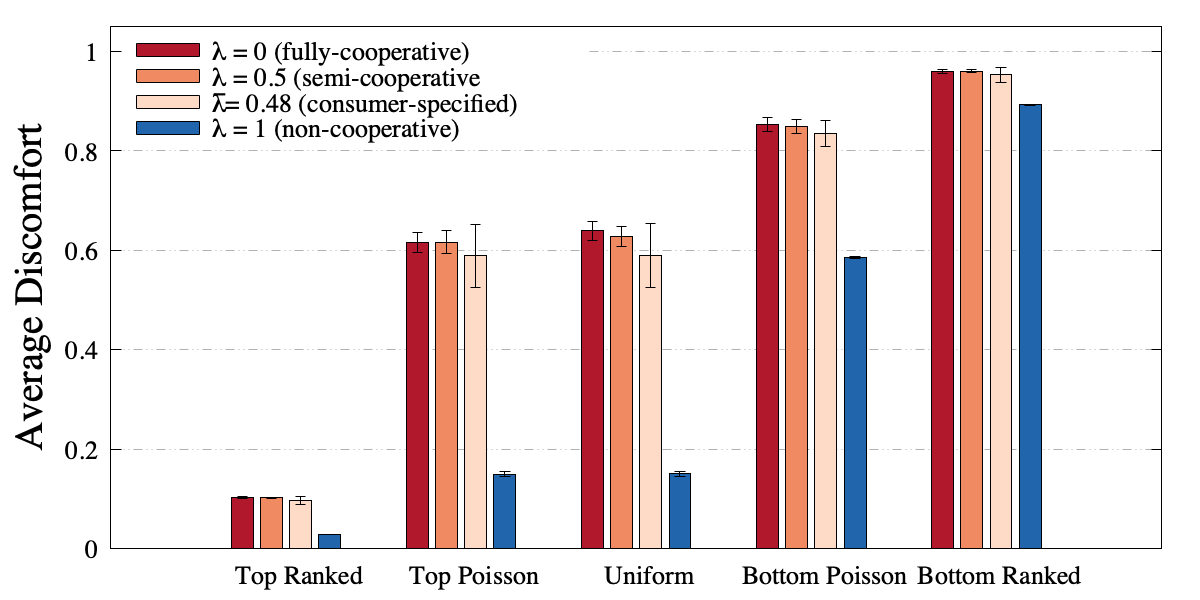}} \\
\subfloat[Unfairness (Equation~\ref{eq:unfairness}) across consumers' selected plans]{\includegraphics[width = 0.45\textwidth]{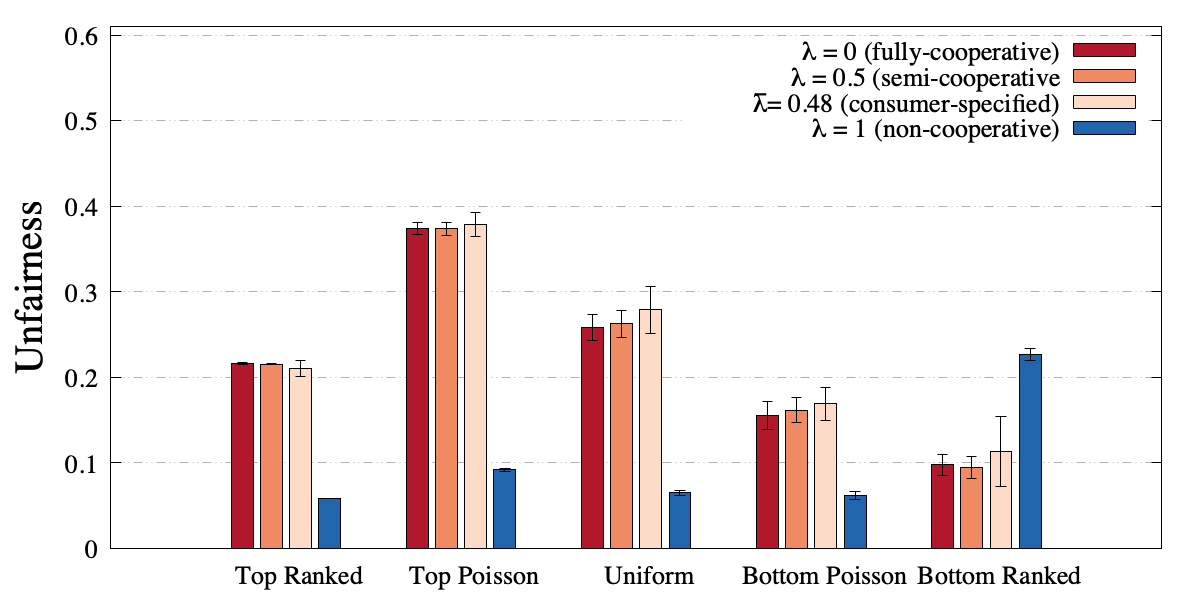}} \\
\caption{\hlc{Total demand variance, average discomfort,} and unfairness across various plan sampling mechanisms for different values of $\lambda$. 
The average $\lambda$ specified by consumers is 0.48. 
The $\lambda$ = 0, 0.5, and 1 are set as system-wide parameters, meaning all agents have the same value of $\lambda$.}
\label{fig:IEPOSRes}
\end{figure}
%%%%
%%%% 
%%%%
\begin{figure*}[!htb]
	\centering
	 \includegraphics[width=0.93\textwidth]{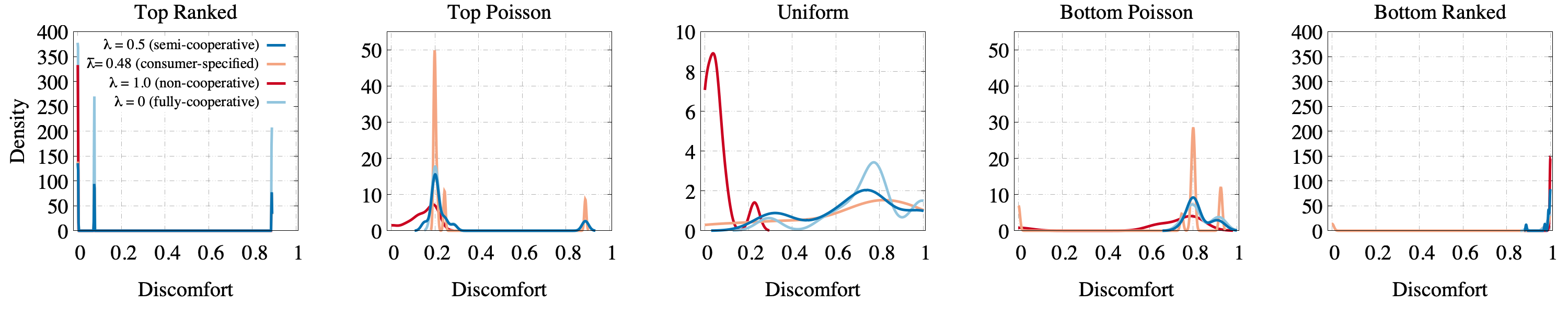}s
	 \caption{Density of consumers' \hlc{discomfort} across different plan sampling mechanisms.}
	\label{fig:LCKDE}
	\end{figure*}
%%%%
%%%%
%%%%
\begin{figure*}[!htb]
	\centering
	 \includegraphics[width=0.93\textwidth]{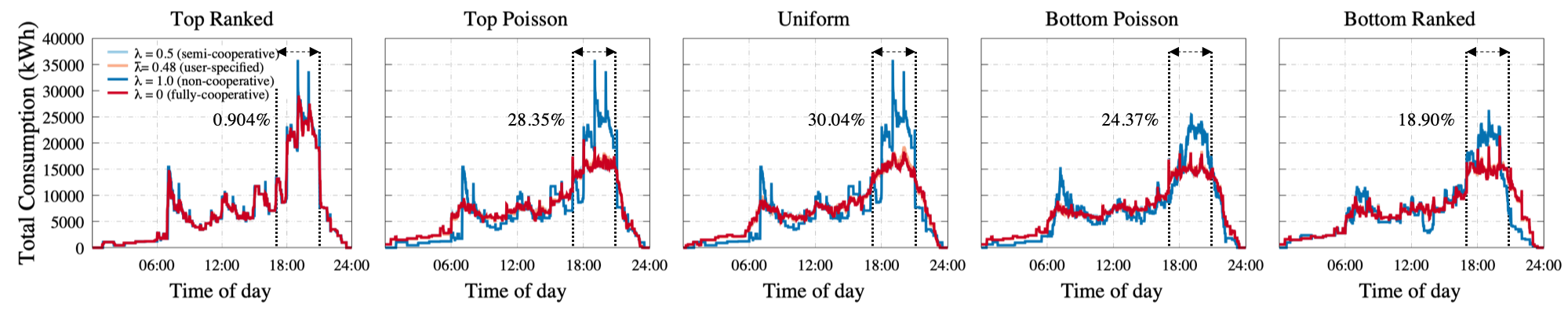}
	 \caption{\hlc{Consumers' total energy demand and peak-time load-shifting} for different values of $\lambda$ across different plan sampling mechanisms.
	 \hlc{The collective goal is to reduce demand peaks by minimizing the variance to consumers' total energy demand.
	 Hence, the more "flat" the energy consumption, the lower the demand variance.
	 The vertical lines indicate peak demand time between 5-9pm, and the adjacent percentages illustrate the peak-time load-shifting capability of the framework; calculated as the percentage of the baseline demand ($\lambda=1$) that can be shifted away from peak hours by leveraging consumer cooperation ($\lambda=0$).}}
	\label{fig:consOverTime}
	\end{figure*}
%%%%
%%%%
%%%%
\begin{figure*}[!htb]
	\centering
	 \includegraphics[width=0.93\textwidth]{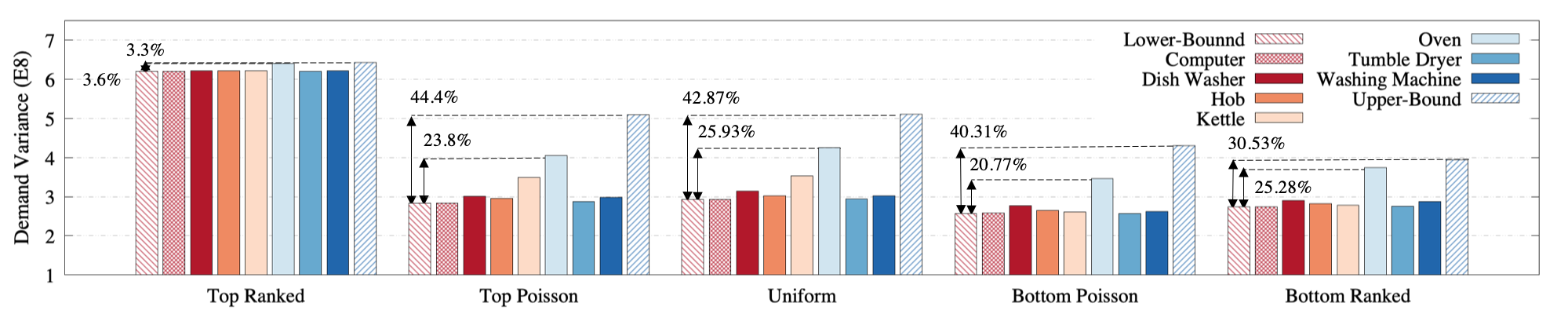}
	 \caption{Impact of individual appliances on \hlc{reducing demand peaks}.
	 The ``Computer" illustrates the scenario where the flexibility of computer schedules are set to 0.
	 By doing so, the computer is effectively excluded from the demand response program, and so on.
	 The ``lower-bound" is the scenario where all of the appliances are included in the demand response program, with schedule flexibility determined by the consumer. 
	 The ``upper-bound" scenario is when the schedule flexibility of all appliances are set to 0.}
	\label{fig:applianceExc}
	\end{figure*} 
%%%%
%%%%
%%%%
\begin{table*}[!htb]
\centering
\caption{\hlc{Determining factors in appliance impact on reducing demand peaks:
The appliances are sorted and ranked according to various usage factors.
The total usage duration refers to scenarios where the given appliance is scheduled multiple times per day.
Additionally, the ranking based on the impact on reducing demand peaks is illustrated.
The correlation between each factor and the impact ranking is calculated based on Kendall ranked correlation coefficient}~\cite{spearman1961proof}.}
	\scriptsize
	\resizebox{\textwidth}{!} {%
\begin{tabular}{llllllll} 
           \toprule
\textbf{Factors} (Correlation to Impact Ranking) & \textbf{Rank 1} & \textbf{Rank 2} & \textbf{Rank 3} & \textbf{Rank 4} & \textbf{Rank 5} & \textbf{Rank 6} & \textbf{Rank 7} \\
          \midrule
\textbf{Flexibility} (-0.14) & Hob (60') & Tumble Dryer (60') & {Oven (58')} & Washing Machine (50') & Computer (47')& Dish Washer (38') & Kettle (34')  \\
          \addlinespace
\textbf{Avg. Duration per Use} (-0.52) & Computer (300') & Washing Machine (101') & {Tumble Dryer (78')} & Oven (52') & Dish Washer (46')& Hob (37') & Kettle (16')  \\
          \addlinespace
\textbf{Avg. Usage Duration per Day} (-0.04) & Computer (4650') & Washing Machine (2077') & {Oven (1269')} & Dish Washer (463') & Kettle (453')& Hob (407') & Tumble Dryer (293')  \\
          \addlinespace
\textbf{Relative Flexibility} (0.52) & Hob (2.22) & Kettle (1.45)& {Oven (1.33)} & Dish Washer (1.03)& Tumble Dryer (0.78)& Washing Machine (0.51)& Computer (0.16)\\
          \addlinespace
\textbf{Energy Consumption kWh} (0.61) & {Oven (3000)} & Tumble Dryer (2073) & Kettle (1992) & Dish Washer (1018) & Hob (1000) & Washing Machine (524) & Computer (27)\\
          \addlinespace
\textbf{Percentage of Plans in Dataset} (0.33) & {Oven (23.90\%)} & Washing Machine (19.52\%)& Kettle (19.04\%)& Computer (14.76\%)& Hob (10.47\%)& Dish Washer (9.52\%)& Tumble Dryer (3.57\%)\\
          \midrule
\textbf{Impact on Reducing Demand Peaks} & {Oven} & Kettle & Dish Washer & Hob & Washing Machine & Tumble Dryer & Computer \\
          \bottomrule
\end{tabular}
}
\label{T:applianceRank} 
\end{table*}
%%%%
%%%%
%%%%
\begin{figure}[!htb]
	\centering
	 \includegraphics[width=0.45\textwidth]{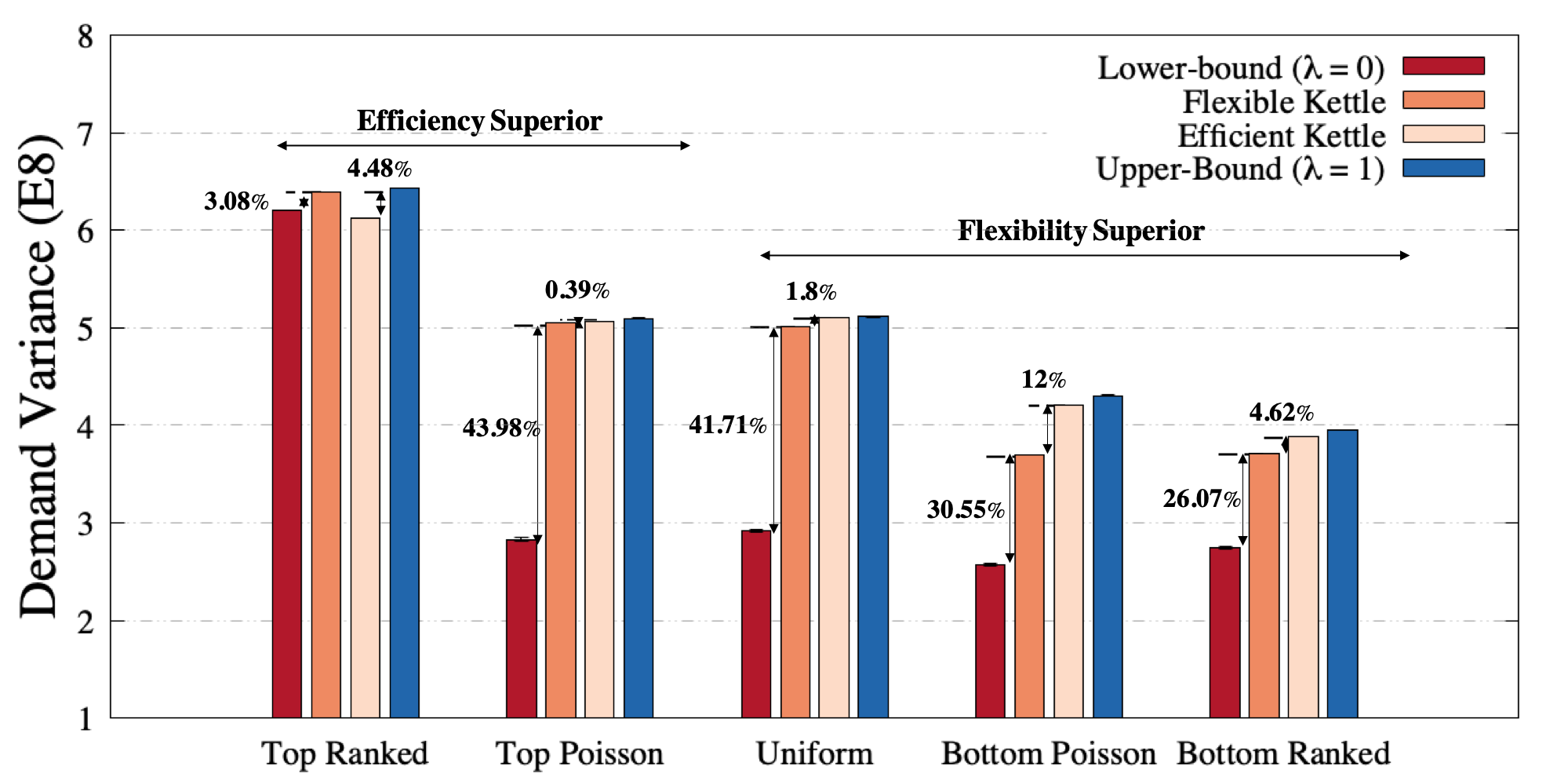}
	 \caption{Comparison of the \hlc{demand variance} between the efficient kettle and the flexible kettle. 
	 Energy consumption of the efficient kettle is based on estimated reduce in energy consumption if the consumers do not overfill\cite{murray2016understanding}. 
	 The flexible kettle is the scenario where only the kettle is flexible and the schedule flexibility of all other appliances is set to 0. 
	 The lower-bound is the scenario where appliances are flexible, and upper-bound is where non of the appliance are flexible.}
	\label{fig:consOverTime-Kettle}
	\end{figure}	
%%%%
%%%%
%%%%
\begin{figure*}[!htb]
	\centering
	 \includegraphics[width=0.93\textwidth]{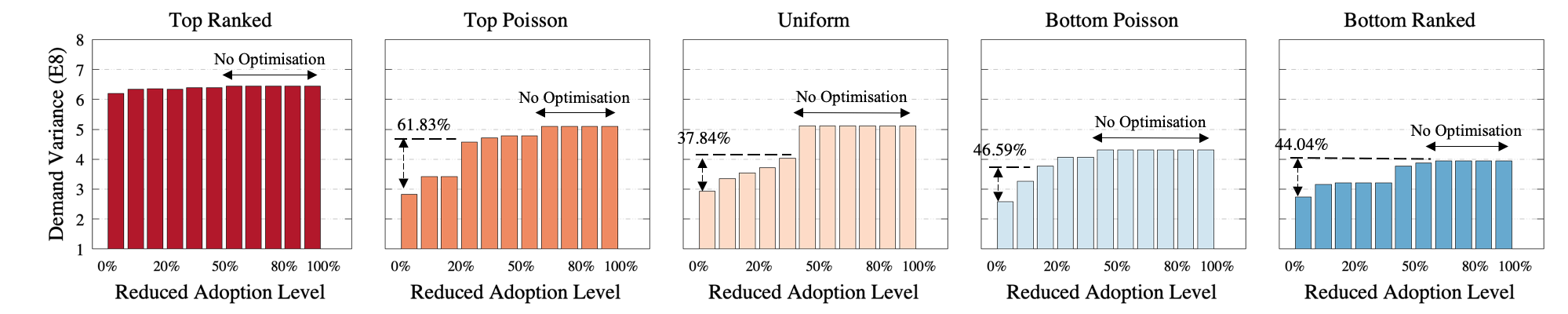}
	 \caption{Impact of reduced levels of participation on the demand response program. 
	 The consumers are sorted in a descending order according to their $\lambda$ value.
	 The $n\%$ reduced participation means that the top $n$ percent of consumers with $\lambda \neq 1$ is changed to 1.
	 \hlc{The ``no optimization" regions indicate the scenarios where, due to the lack of consumers' flexibility, no reduction of demand variance is possible.}}
	\label{fig:varParticipation}
	\end{figure*}
%%%%
%%%%
%%%%
\subsection{Experimental Results}
\label{S:IEPOSRes}
%%%%
%%%%
%%%%
This section illustrates the results of the experiments based on the methodology described in Section~\ref{S:Methodology}.
%%%%
%%%%
%%%%
\subsubsection{Flexible Appliance Scheduling \& Reducing Demand Peaks}
\label{S:globalCost}
Figure~\ref{fig:IEPOSRes}a illustrates the \hlc{variance of consumers' total demand} at the final iteration of I-EPOS, for different sampling mechanisms and across different values of $\lambda$. 
\hlc{Note that having lower variance means the demand is more spread out across the day, resulting in reduced demand peaks.}
The general trend across all sampling mechanisms is that with the increase of $\lambda$, the \hlc{demand variance increases, as agents with higher $\lambda$ values focus more minimizing discomfort than cooperating to reduce demand peaks} (Equation~\ref{eq:IEPOSExpanded}).
The Top Poisson is the most sensitive \hlc{case, with 44.48\% increase in variance from $\lambda=0$ to $\lambda=1$.} 
More drastic is the impact of different sampling mechanisms on \hlc{demand variance}.
\hlc{For instance,} changing the plan sampling mechanism from Top Ranked to Top Poisson, \hlc{decreases demand variance} by 54.35\% for $\lambda=0$, 54.29\% for $\lambda=0.5$, 52.44\% for consumer specified, and 20.80\% for $\lambda=1$. 
These differences are due to the entropy and diversity of consumers' sampled plans~\cite{pournaras2017self,pournaras2014decentralized}.
\hlc{For instance}, in Top Ranked the agents always sample 10 plans with the lowest \hlc{discomfort; thus, the plans are very similar and the entropy among them is low.}
Hence, I-EPOS cannot perform effective optimization in such a non-diverse plan space. 
Overall, the best performing sampling mechanism is the Bottom Poisson that reduces the \hlc{demand variance by 51.71\% compared} to Top Ranked.
%%%%
%%%%
%%%%
\subsubsection*{\textbf{Discomfort \& Unfairness}}
\label{S:avgLocalcost}
Figure~\ref{fig:IEPOSRes}b illustrates the average \hlc{discomfort among consumers for different sampling mechanisms and across different values of $\lambda$.}
Within the same plan sampling mechanism, \hlc{higher $\lambda$ values result in lower average discomfort.}
This is because the higher $\lambda$ values correspond to \hlc{less} cooperative consumers, \hlc{which tend to choose the plans with the lower discomfort.}
\hlc{Hence,} the best performing sampling mechanism regarding reducing demand peaks (Bottom Poisson), results in one of the highest average \hlc{discomfort}.
\hlc{This phenomenon is further studied in Figure}~\ref{fig:LCKDE}, \hlc{where the distribution of discomfort across selected plans at the final iteration of I-EPOS is illustrated.}
For instance, in Top Ranked, the discomfort values are highly concentrated around 0, while in the Bottom Ranked the values are concentrated around 1\footnote{This observation supports the insight regarding differences in plan entropy and diversity across plan sampling mechanisms (Section~\ref{S:globalCost}).}.
\hlc{Overall,} by changing the plan sampling mechanism from Top Ranked to Top Poisson, Uniform, Bottom Poisson, and Bottom Ranked, the average \hlc{discomfort} rises by a factor of 5.9, 0.5, 1.05, and 0.42, respectively.
Figure~\ref{fig:IEPOSRes}c shows the unfairness calculated by Equation~\ref{eq:unfairness}. 
With the exception of the Bottom Ranked sampling mechanism, the unfairness decreases with the increase of $\lambda$.
Among the plan sampling mechanism, the average unfairness in Top Ranked, Top Poisson, Uniform, Bottom Poisson, and Bottom Ranked are 0.175, 0.304, 0.216, 0.137, and 0.133, respectively.
%%%%
%%%%
%%%%
\subsubsection*{\textbf{Peak-time Load-shifting}}
\label{S:totalEnergyDemand}
Figure~\ref{fig:consOverTime} shows consumers' aggregated energy demand across different sampling mechanisms. 
\hlc{Note that while the total energy demand is the same in all scenarios, by utilizing the framework, the demand is distributed more evenly across the day.}
Additionally, Figure~\ref{fig:consOverTime} \hlc{illustrates the peak demand reduction capability of the framework, calculated as the percentage of the baseline demand ($\lambda=1$) that can be shifted away from peak hours (5-9pm) by leveraging consumer flexibility ($\lambda=0$).}
%%%%
%%%%
%%%%
\subsubsection{Impact of Individual Appliances on Reducing Demand Peaks}
\label{S:applianceImpact}
Figure~\ref{fig:applianceExc} illustrates the impact of various appliances on \hlc{reducing demand peaks.}
\hlc{Overall,} across all plan sampling mechanisms, the oven has the highest impact.
After the over, the kettle and dish washer are the next appliances with the highest impact.
This impact is attributed to a multitude of factors, such as appliance scheduling flexibility, \hlc{average duration per usage, average usage duration per day}, relative flexibility (Table~\ref{T:flexDay}), energy consumption, and the number of plans in the dataset.
Table~\ref{T:applianceRank} ranks the appliances based on these factors, \hlc{and shows the correlation of each factor with the impact ranking.
Note that while no single factor can fully explain the impact ranking of various appliances, relative flexibility (0.52) and energy consumption (0.61) have the highest positive correlation to the impact ranking, while average usage duration is negatively correlated\footnote{As the usage of appliances with high usage duration, such as Computer, cannot be effectively shifted in time.}, and the correlation with flexibility is not significant.
 This indicates that to achieve effective reduction in demand peaks, the demand response programs can focus on appliances with high relative flexibility and high energy consumption.
Table}~\ref{T:applianceExc} \hlc{in Appendix B illustrates the detailed numerical results of this experiment.}
%%%%
%%%%
%%%%
\subsubsection{Increased Efficiency vs Flexible Coordinated Scheduling}
\label{S:efficiencyVSflexibility}
Figure~\ref{fig:consOverTime-Kettle} illustrates the results of scenarios (a) ``Efficient Kettle" and (b) ``Flexible Kettle", defined in Section~\ref{S:M:IEOFS}.
\hlc{In these experiments,} the upper-bound is the scenario where none of the appliances are flexible, and the lower-bound is where all appliances are flexible.
In scenario (a), the total energy demand decreases by 4.73\%, and on average the \hlc{demand variance} reduces by 2.04\% compared to the upper-bound, \hlc{indicating that more efficient usage of kettles indeed reduces demand peaks}.
In scenario (b), the total energy demand remains the same. 
However, on average the \hlc{demand variance} reduces by 29.8\% compared to the upper-bound. 
\hlc{A} critical observation here is the role of plan sampling mechanism. 
In the schemes with lower consumer flexibility, i.e., Top Ranked, Top Poisson (referred to as ``Efficiency Superior"), the increased efficiency approach performs better.
However, in schemes where consumers are relatively more flexible, i.e., the Uniform, Bottom Poisson, and Bottom Ranked  (referred to as ``Flexibility Superior"), optimized flexible scheduling performs better.
 These results show that in scenarios with high overall flexibility and cooperation, flexible coordinated scheduling of appliances can further contribute to the effective \hlc{reduction of demand peaks}.
%%%%
%%%%
%%%%
\subsubsection{Variable Adoption of Recommended Plans}
\label{S:varParticipation}
Figure~\ref{fig:varParticipation} illustrates the increase in \hlc{demand variance}, due to the reduced adoption of recommended plans by the consumers.
Top Poisson is the most sensitive sampling mechanism, as with 30\% reduced participation, the \hlc{demand variance} already increases by 61.83\%.
The Bottom Ranked is the most resilient sampling mechanism where the \hlc{demand variance} can be reduced by 16.98\% even if 40\% of consumers do not adopt the recommended plan.
%%%%
%%%%
%%%%
\section*{Summary of the Findings}\label{sec:findings}
The key findings of this paper are summarized as follows:
\begin{itemize}[leftmargin=*]
\item Consumers' flexibility in appliance scheduling depends on various socio-technical factors, such as the appliance type, usage habits, and the time of the day.
\item \hlc{Cooperative consumers contribute energy consumption plans with higher levels of entropy and diversity, and also provide more flexibility to the demand response program, thus allowing for more effective reduction of demand peaks.}
\item Among the 7 appliances involved in the experiments, the oven has the most significant role in \hlc{reducing demand peaks.
Among appliance characteristics, higher relative flexibility and energy consumption are the two factors with the highest correlation with the impact on reducing demand peaks.}
\item \hlc{While increased appliance efficiency is an effective approach to improve the grid reliability, in comparison, flexible coordinated scheduling can further reduce demand peaks, especially in scenarios with an overall high consumer flexibility.}
\item Decrease in consumer participation and adoption \hlc{of recommended plans negatively affect the collective goal of reducing demand peaks.}
However, the degree of such effect varies among the plan sampling mechanisms, with Bottom Ranked being the most resilient.
\end{itemize}
%%%%
%%%%
%%%%
\section{Conclusion and Future Work}
\label{S:conclusion}
\hlc{This paper concludes that socio-technical Smart Grid optimization aiming to reduce demand peaks, is feasible by decentralized coordination of flexible appliance-level energy usage schedules.
The findings of this paper confirm that consumers' flexibility in appliance usage indeed varies across several socio-technical factors, such as applicant type, usage habits, and time of the day.
Additionally, consumers cooperation and willingness to sacrifice comfort (by voluntary contribution of flexibility) to improve Smart Grid stability and reliability, greatly facilitates the reducing of demand peaks across the day, with the oven having the highest system-wide potential for this.}
Further research can address the inclusion of the markets and the role of dynamic energy pricing, such as incentive mechanisms (e.g., discounts).
The proposed framework can be evaluated in pilot tests with energy utility companies to analyze the long-term improvement and efficiency.
Moreover, the framework and methodology of this paper can be expanded to study additional appliances, such as PV panels, electric heating, and cooling devices, where power consumption can be adjusted as well.
Lastly, additional societal and behavioral factors, such as consumers' environmental awareness, and carbon emissions in appliance usage, can be included in the proposed framework to provide a more comprehensive study.
%%%%
%%%%
%%%%
\bibliographystyle{model1-num-names}
\bibliography{reference.bib}
%%%%
%%%%
%%%%
\begin{figure}
	\centering
	 \includegraphics[width=0.6\linewidth]{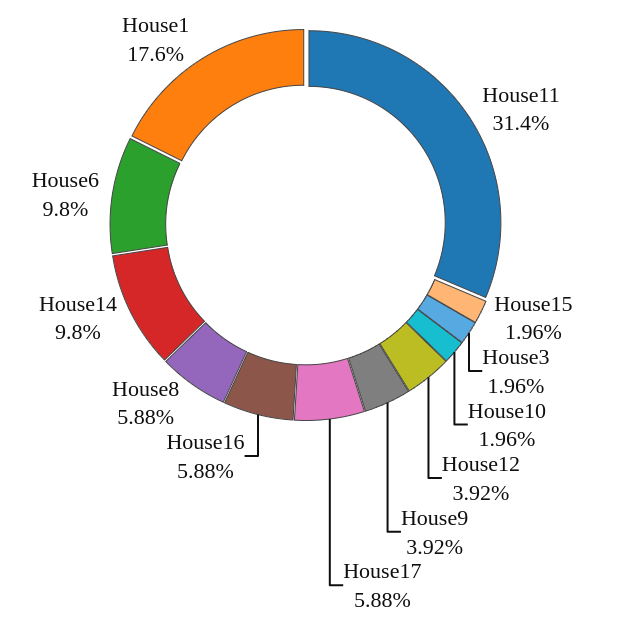}
	 \caption{Distribution of assigned houses across all consumers. 
	 The profiles are based on Table~\ref{T:REFITProfiles}.}
	\label{fig:ProfDist}
	\end{figure} 
%%%%
%%%%
%%%%
\begin{table}[!htb]
\centering
\caption{Households in the REFIT dataset.}
\label{T:REFITProfiles}
\scriptsize
\begin{tabular}{llllll} 
\toprule
\textbf{House} & \textbf{Occupancy} & \textbf{Year Built} & \textbf{Appliances} & \textbf{Type} & \textbf{Size} \\
\midrule
1 & 2 & 1975-1980 & 35 & Detached & 4 Beds \\
\addlinespace
2 & 4 & - & 15 & Semi-Detached & 4 Beds \\
\addlinespace
3 & 2 & 1988 & 27 & Detached & 3 Beds \\
\addlinespace
4 & 2 & 1850-1899 & 33 & Detached & 3 Beds \\
\addlinespace
5 & 4 & 1878 & 44 & Mid-Terrace & 4 Beds \\
\addlinespace
6 & 2 & 2005 & 49 & Detached & 4 Beds \\
\addlinespace
7 & 4 & 1965-1974 & 25 & Detached & 4 Beds \\
\addlinespace
8 & 2 & 1966 & 35 & Detached & 3 Beds \\
\addlinespace
9 & 2 & 1919-1944 & 24 & Detached & 2 Beds \\
\addlinespace
10 & 4 & 1919-1944 & 31 & Detached & 3 Beds \\
\addlinespace
11 & 1 & 1945-1963 & 25 & Detached & 3 Beds \\
\addlinespace
12 & 3 & 1991-1995 & 26 & Detached & 3 Beds \\
\addlinespace
13 & 4 & Post 2002 & 28 & Detached & 4 Beds \\
\addlinespace
14 & 1 & 1965-1974 & 19 & Semi-Detached & 3 Beds \\
\addlinespace
15 & 6 & 1981-1990 & 48 & Detached & 5 Beds \\
\addlinespace
16 & 3 & Mid 60s & 22 & Detached & 3 Beds \\
\addlinespace
17 & 2 & 1965-1974 & 34 & Detached & 3 Beds \\
\addlinespace
18 & 4 & 1945-1964 & 26 & Semi-Detached & 3 Beds \\
\addlinespace
19 & 2 & 1965-1975 & 39 & Detached & 3 Beds \\
\addlinespace
20 & 4 & 1981-1990 & 23 & Detached & 3 Beds \\
\bottomrule
\end{tabular}
\end{table}
%%%%
%%%%
%%%%
\begin{table}[!htb]
\centering
\caption{Appliance energy consumption (kWh) for households in the REFIT dataset. TD: Tumble Dryer, WD: Tasing machine, DW: Dish Washer.}
\label{T:REFITHouses}
\scriptsize
\begin{tabular}{llllll}
\toprule
\textbf{House / Appliance} & \textbf{TD} & \textbf{WM} & \textbf{Computer} & \textbf{DW} & \textbf{Kettle}\\ 
\midrule
 House 1   & 472          & 513             & 29       & 1379        & -      \\
 House 2   & -            & 327             & -        & 770         & 2257   \\
 House 3   & 1373         & 492             & 16       & 1150        & 1550   \\
 House 4   & -            & 56              & 52       & -           & 1703   \\
 House 5   & 766          & 66              & 66       & 182         & 2352   \\
 House 6   & -            & 369             & 66       & 778         & 2192   \\
 House 7   & 2075         & 442             & -        & 613         & 1913   \\
 House 8   & -            & 273             & 19       & -           & 2340   \\
 House 9   & -            & 507             & -        & 700         & 2359   \\
 House 10  & -            & 349             & -        & 364         & -      \\
 House 11  & -            & 91              & 10       & 753         & 1841   \\
 House 12  & -            & -               & -        & -           & 2482   \\
 House 13  & 152          & 203             & 39       & 1250        & 1542   \\
 House 14  & 1476         & 495             & 20       & 532         & 2521   \\
 House 15  & -            & 300             & 27       & 1239        & -      \\
 House 16  & 1594         & 373             & 20       & -           & 1689   \\
 House 17  & -            & 377             & 26       & 1021        & -      \\
 House 18  & -            & 161             & -        & -           & 2448   \\
 House 19  & 1097         & 293             & 75       & 434         & 2350   \\
 House 20  & 1240         & 434             & -        & 350         & 1276 \\
\bottomrule
\end{tabular}
\end{table}
%%%%
%%%% 
%%%%
\begin{table}[!htb]
\centering
\caption{Distribution of appliance schedules throughout the day.
The morning hours are between 00:00 - 08:59, mid day between 09:00 - 16:59, and evening between 17:00 - 23:59. 
This splitting is made based on the common demand patters throughout the day\cite{richardson2010domestic}}
\label{T:scheduleDay}
\scriptsize
\begin{tabular}{llll} 
\toprule
\textbf{Appliance} & \textbf{Morning (\%)} & \textbf{Mid Day (\%)} & \textbf{Evening (\%)} \\
\midrule
Computer & 16 (25.80\%) & 26 (41.93\%) & 20 (32.25\%)  \\
\addlinespace
Dish Washer & 2 (5\%) & 3 (7.5\%) & 35 (87.5\%)  \\
\addlinespace
Hob & 2 (4.54\%) & 18 (40.9\%) & 24 (54.54\%) \\
\addlinespace
Kettle & 37 (46.25\%) & 18 (22.5\%) & 25 (31.25\%)  \\
\addlinespace
Oven & 27 (27.83\%) & 19 (19.58\%) & 51 (52.57\%)  \\
\addlinespace
Tumble Dryer & 2 (13.33\%) & 10 (66.66\%) & 3 (20\%) \\
\addlinespace
Washing Machine & 12 (14.63\%) & 21 (25.60\%) & 49 (59.75\%)  \\
\bottomrule
\end{tabular}
\end{table} 
%%%%
%%%%
%%%%
\begin{figure}[t]
\subfloat[\hlc{Variance of consumers' total demand across the day}]{\includegraphics[width = 0.45\textwidth]{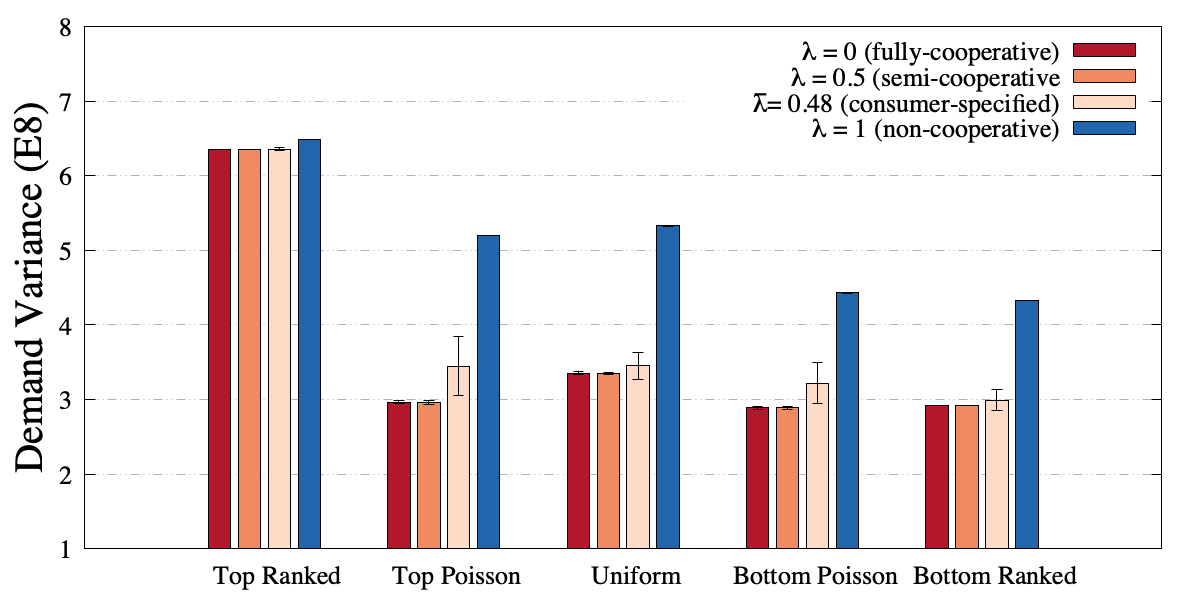}} \\ 
\subfloat[\hlc{Average discomfort }(Equation~\ref{AVGDiscomfort}) of consumers' selected plans]{\includegraphics[width = 0.45\textwidth]{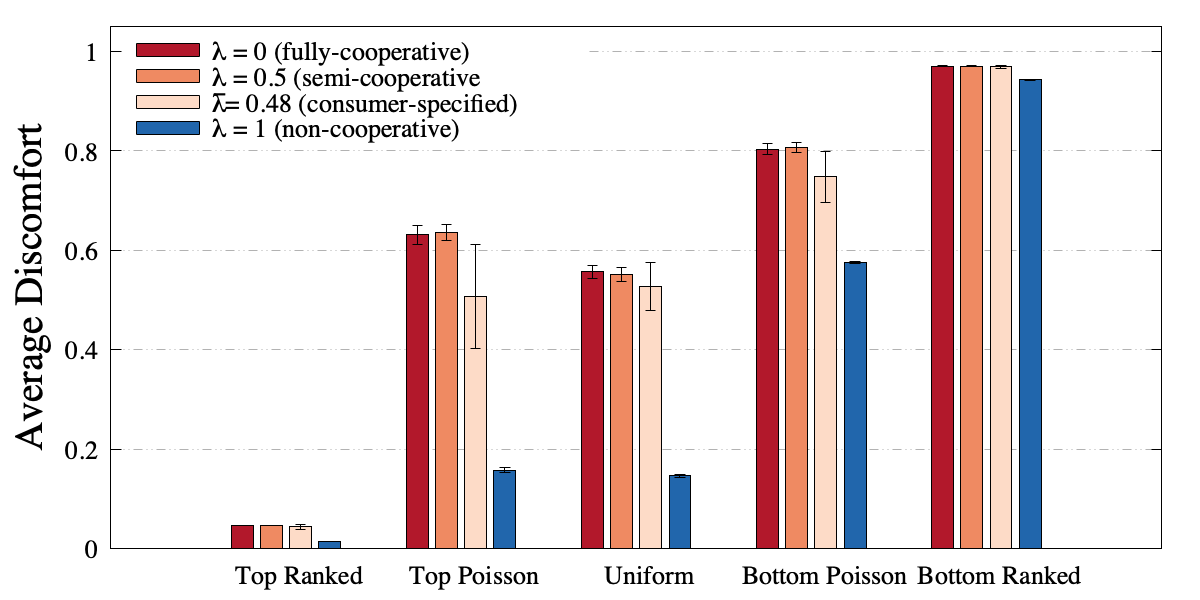}} \\
\subfloat[Unfairness (Equation~\ref{eq:unfairness}) across consumers' selected plans]{\includegraphics[width = 0.45\textwidth]{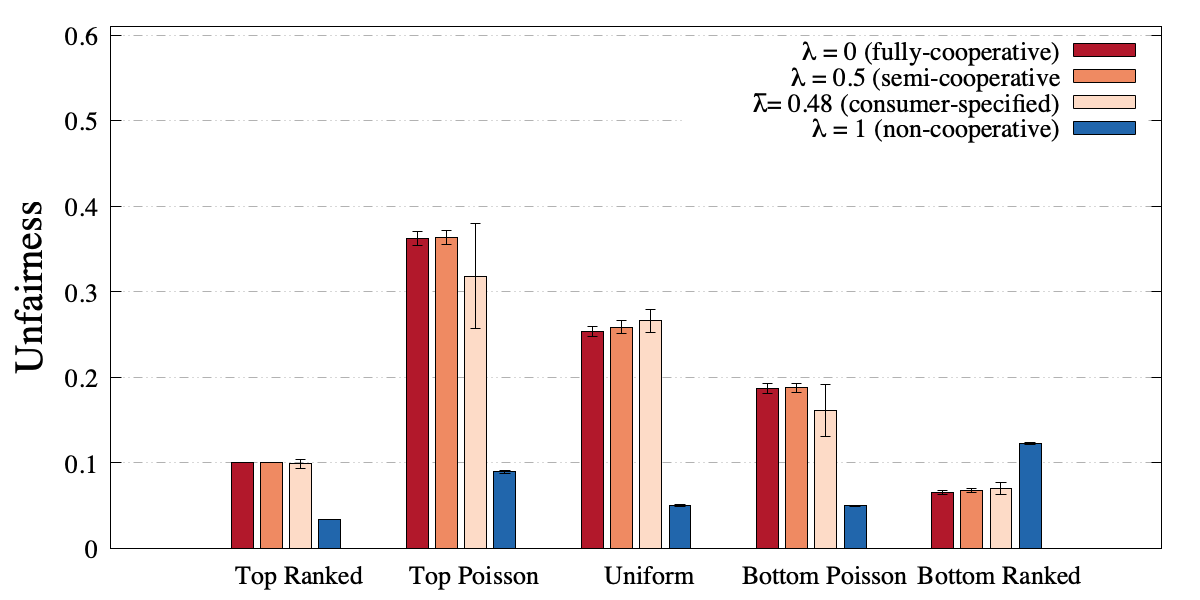}} \\
\caption{\hlc{Total demand variance, average discomfort,} and unfairness for 5 plans per agent across various plan sampling mechanisms for different values of $\lambda$. 
The average $\lambda$ specified by consumers is 0.48. 
The $\lambda$ = 0, 0.5, and 1 are set as system-wide parameters, meaning all nodes have the same value of $\lambda$.}
\label{fig:IEPOSRes5}
\end{figure}
%%%%
%%%% 
%%%%
\begin{figure}[t]
\subfloat[\hlc{Variance of consumers' total demand across the day}]{\includegraphics[width = 0.45\textwidth]{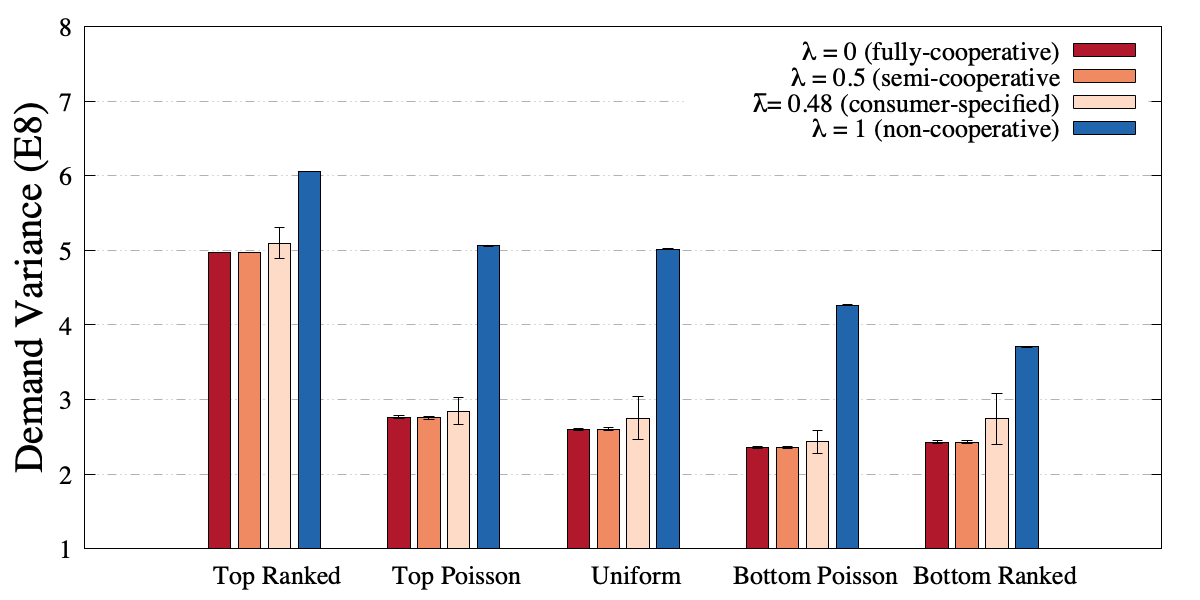}} \\ 
\subfloat[\hlc{Average discomfort }(Equation~\ref{AVGDiscomfort}) of consumers' selected plans]{\includegraphics[width = 0.45\textwidth]{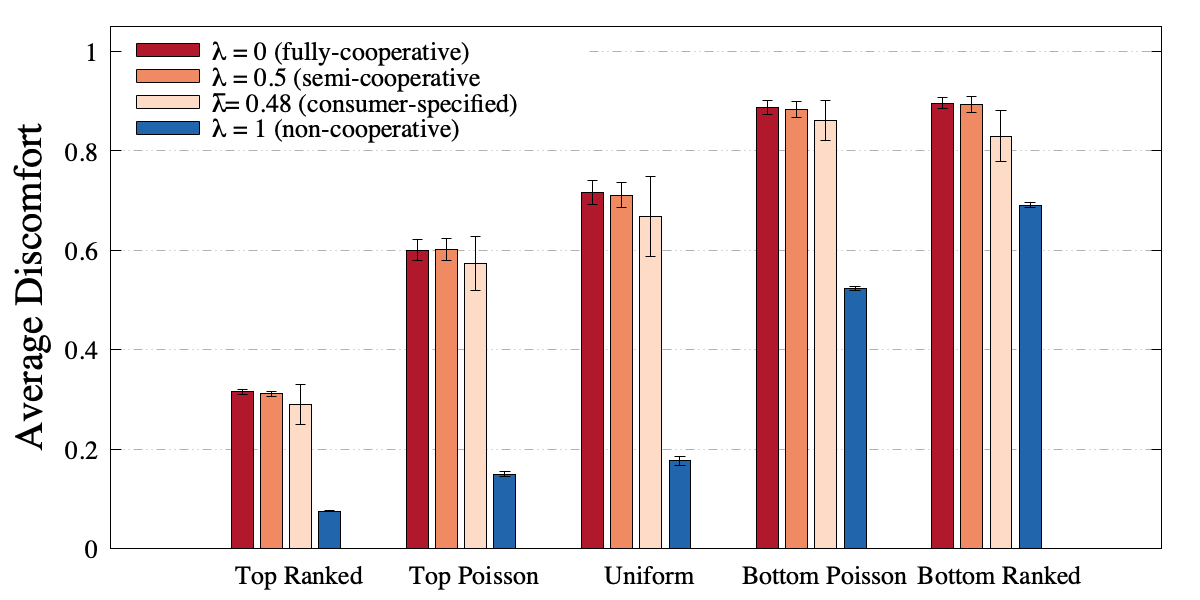}} \\
\subfloat[Unfairness (Equation~\ref{eq:unfairness}) across consumers' selected plans.]{\includegraphics[width = 0.45\textwidth]{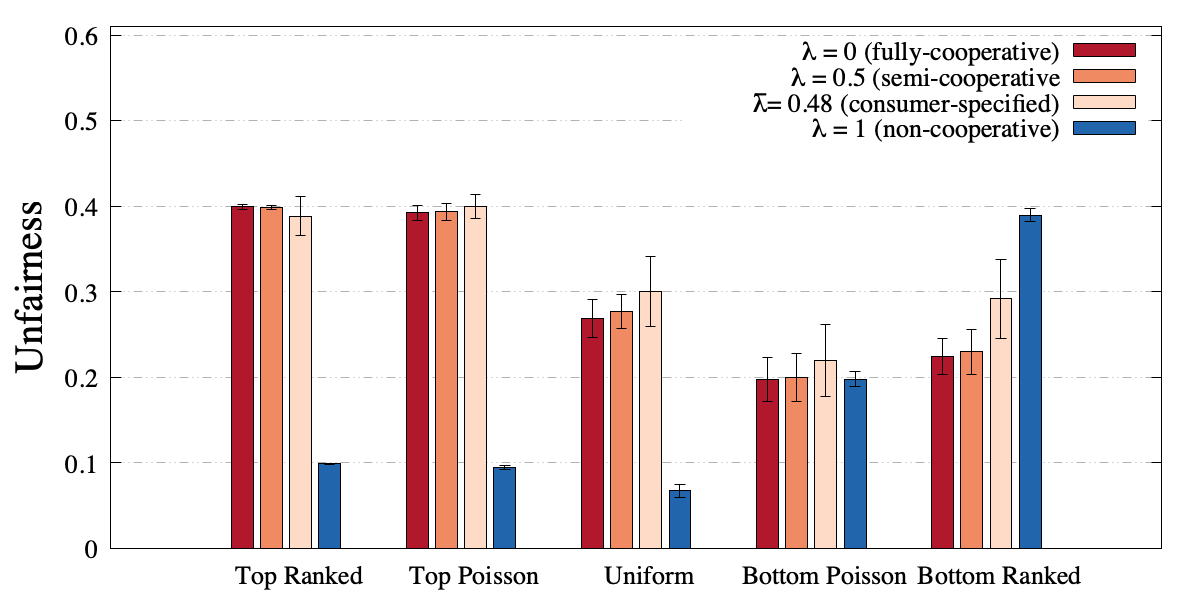}} \\
\caption{\hlc{Total demand variance, average discomfort,} and unfairness for 100 plans per agent across various plan sampling mechanisms for different values of $\lambda$. 
The average $\lambda$ specified by consumers is 0.48. 
The $\lambda$ = 0, 0.5, and 1 are set as system-wide parameters, meaning all nodes have the same value of $\lambda$.}
\label{fig:IEPOSRes100}
\end{figure}
%%%%
%%%% 
%%%%
\begin{table*}[t]
\centering
\caption{Effect of individual appliances on \hlc{demand variance}.
The ``Computer" illustrates the scenario where the flexibility of computer schedules are set to 0.
By doing so, the computer is effectively excluded from the demand response program, and so on.
The ``lower-bound" is the scenario where all of the appliances are included in the demand response program, with schedule flexibility determined by the consumer. 
The ``upper-bound" scenario is when the schedule flexibility of all appliances are set to 0.}
	\scriptsize
\begin{tabular}{llllllllll} 
           \toprule
\textbf{\hlc{Demand Variance} (E8)} & \textbf{Lower-Bound} & \textbf{Computer} & \textbf{Dish Washer} & \textbf{Kettle} & \textbf{Hob} & \textbf{Oven} & \textbf{Tumble Dryer} & \textbf{Washing Machine} & \textbf{Upper-Bound} \\
          \midrule
Top Ranked & 6.1971 & 6.1976 & 6.2165 & 6.2223 & 6.2206 & 6.4095 & 6.2000 & 6.2169 & 6.4346 \\
          \addlinespace
Top Poisson & 2.8289 & 2.8289 & 3.006 & 2.9551 & 3.4916 & 4.0458 & 2.8759 & 2.9771 & 5.0961 \\
          \addlinespace
Uniform & 2.9202 & 2.9231 & 3.1401 & 3.0200 & 3.5315 & 4.2458 & 2.9424 & 3.0255 & 5.1120 \\
          \addlinespace
Bottom Poisson & 2.5694 & 2.5737 & 2.7656 & 2.6455 & 2.6049 & 3.4636 & 2.5707 & 2.6234 & 4.3049 \\
          \addlinespace
Bottom Ranked & 2.7426 & 2.7446 & 2.9027 & 2.8246 & 2.7845 & 3.7413 & 2.7479 & 2.8780 & 3.9483 \\
          \bottomrule
\end{tabular}
\label{T:applianceExc}
\end{table*}
%%%%
%%%%
%%%%
\appendix
\section{A: Profile Assignment to consumers}
\label{A:profileAssignment}
%%%%
%%%%
%%%%
The estimation of consumers' appliance energy consumption is based on the households from REFIT, illustrated in Table~\ref{T:REFITProfiles}. 
Based on consumers' household information, each consumer is assigned to one of the REFIT households.
The matching is performed using a linear scoring function: 
%%%%
%%%%
%%%%
\begin{equation}
\label{eq:profileMatching}
	\begin{aligned}
	&Score = 0.533*Occupancy~+~0.267*Size \\
	&+~0.133*Type~+~0.067*\text{\textit{Year-built}}
	\end{aligned}
\end{equation}
%%%%
%%%%
%%%%
The weights are based on the importance each feature on determining the household.
The linear combination is used for the sake of simplicity and interpretability. 
Though, more complex formulations are possible.
For each consumer, the matching scores are calculated for all houses. 
The house with the highest matching score is assigned to the consumer. 
The energy consumption of consumers' appliances are then calculated based on the assigned house.
This consumption data is illustrated in Table~\ref{T:REFITHouses}.
If the assigned house does not include one of the consumers' appliances, the next best-matched house is used to calculate the appliance energy consumption.
The energy consumption of the oven and hob are based on average consumption of some models available in the market.
Figure~\ref{fig:ProfDist} illustrates the consumers' assigned house distribution in the collected dataset. 
%%%%
%%%%
%%%%
\section{B: Expanded Experiments and Evaluations}
\label{A:expandedExp}
%%%%
%%%%
%%%%
Table~\ref{T:scheduleDay} shows the distribution of appliance schedules throughout the day. 
The observed usage patterns, such as high usage of Hob around lunch and dinner times, are in accordance with previous research on appliance usage\cite{torriti2017understanding}.
Figure~\ref{fig:IEPOSRes5} and Figure~\ref{fig:IEPOSRes100} illustrate the results of I-EPOS experiment with 5 and 100 plans per agent, respectively. 
The key findings of the results illustrated in the paper, e.g., increase of \hlc{variance} with with increase of $\lambda$, are observed here as well.
The higher number of plans results in lower \hlc{variance}, but higher average discomfort and unfairness.
Table~\ref{T:applianceExc} shows more detailed results of the impact of individual appliances on demand variance.
%%%%
%%%% 
%%%%
\section{C: Detailed Survey}
\label{A:surveyRes}
%%%%
%%%%
%%%%
\clearpage
\onecolumn
%%%%
%%%%
%%%%
\captionof{table}{Survey and results}
\scriptsize
     \begin{tabularx}{\textwidth}{lXXl}
           \toprule
	  \textbf{Category} & \textbf{Question} & \textbf{Answer (\%)} & \textbf{ID}  \\
          \midrule
          \multirow{1}{*}{Demography (D)} & 1. What year were you born? & 1962 (2.17\%) - 1972 (2.17\%) - 1976 (2.17\%) - 1978 (2.17\%) - 1980 (4.35\%) - 1981 (4.35\%) - 1983 (4.35\%) - 1984 (4.35\%) - 1985 (4.35\%) - 1986 (8.70\%) - 1987 (8.70\%) - 1988 (6.52\%) - 1989 (6.52\%) - 1990 (15.22\%) - 1991 (13.04\%) - 1994 (4.35\%) - 1995 (4.35\%) - 1996 (2.17\%) & D.1 \\
          \addlinespace
                    & 2. In which country were you born? & Brazil (6.52\%) - Colombia (2.17\%) - France (4.35\%) - Germany (10.87\%) - Greece (4.35\%) - India (2.17\%) - Indonesia (10.87\%) - Iran (28.26\%) - Italy (4.35\%) - The Netherlands (13.04\%) - Romania (2.17\%) - Russia (2.17\%) - Singapore (2.17\%) - Switzerland (2.17\%) - Turkey (4.35\%) & D.2 \\
                    \addlinespace
                    & 3. What is your gender? & Male (78.26\%) - Female (21.74\%) & D.3 \\
                    \addlinespace
                    & 4. In which country have you lived the longest? & Brazil (6.52\%) - Colombia (2.17\%) - France (4.35\%) - Germany (8.70\%) - Greece (4.35\%) - India (2.17\%) - Indonesia (10.87\%) - Iran (28.26\%) - Italy (4.35\%) - The Netherlands (15.22\%) - Romania (2.17\%) - Russia (2.17\%) - Switzerland (4.35\%) - Turkey (4.35\%) & D.4 \\
                    \addlinespace
                    & 5. What is the highest level of education you have completed? & Short cycle tertiary (2.17\%) - Bachelor or equivalent (13.04\%) - Master or equivalent (73.91\%) - Doctoral or equivalent (10.87\%) & D.5 \\
                    \addlinespace
                    & 6. Which of the following best describes your employment status? & Full time (63.04\%) - Part-time (4.35\%) - Self-employed (2.17\%) - Student (23.91\%) - Unemployed (6.52\%) & D.6 \\
                    \addlinespace
          \cline{1-4}
          \addlinespace
          \multirow{1}{*}{Household (H)} & 1. What type of house do you live in? & Apartment/Flat (63.04\%) - Detached (63.04\%) - Mid-terrace (63.04\%) - Semi-detached (63.04\%) - Other (63.04\%) & H.1 \\
          \addlinespace          
          &2. What is the size of your house? & 1 Bed (28.26\%) - 2 Beds (32.61\%) - 3 Beds (30.43\%) - 4 Beds (4.35\%) - 5 Beds (2.17\%) - 6+ Beds (2.17\%) & H.2 \\
          \addlinespace
          &3. Approximately, when was your house built? & Pre 1900s (8.70\%) - 1920-1929 (6.52\%) - 1930-1939 (4.35\%) - 1950-1959 (2.17\%) - 1960-1969 (17.39\%) - 1970-1979 (6.52\%) - 1980-1989 (15.22\%) - 1990-1999 (10.87\%) - 2000-2009 (15.22\%) - 2010+ (13.04\%) & H.3 \\
         \addlinespace
          &4. What type of house do you live in? & Apartment/Flat (63.04\%) - Detached (63.04\%) - Mid-terrace (63.04\%) - Semi-detached (63.04\%) - Other (63.04\%) & H.4 \\
         \addlinespace
          &5. How many people live in the house? & 1 (26.09\%) - 2 (28.26\%) - 3 (30.43\%) - 4 (10.87\%) - 6+ (4.35\%) & H.5 \\
         \addlinespace
         &6. Which appliance do you have at home? & Washing machine (80.39\%) - Tumble dryer (19.60\%) - Computer-laptop (86.27\%) - Computer-desktop (33.33\%) - Oven (70.50\%) - Hob (23.52\%) - Electric shower (7.84\%) - Dish washer (39.21\%) - Electric heater (29.41\%) - Air conditioner (13.72\%) - Kettle (54.90\%) - Microwave (62.74\%) - Freezer (68.62\%) - Fridge (80.39\%) & H.6 \\
         \addlinespace
         &7. To which devices you would allow automated control for a more efficient energy usage? & Washing machine (82.35\%) - Tumble dryer (27.46\%) - Computer-laptop (25.49\%) - Computer-desktop (11.76\%) - Oven (31.37\%) - Hob (5.88\%) - Electric shower (9.80\%) - Dish washer (47.05\%) - Electric heater (19.60\%) - Air conditioner (17.64\%) - Kettle (15.68\%) - Microwave (23.52\%) - Freezer (47.05\%) - Fridge (56.86\%) & H.7 \\
         \addlinespace
          \cline{1-4}
         \addlinespace
          \multirow{1}{*}{Preferences (P)} & 1. I am concerned about the amount of my residential energy consumption: & 0 (0\%) - 1 (13.46\%) - 2 (17.31\%) - 3 (46.15\%) - 4 (23.08\%) & P.1 \\
          \addlinespace          
          &2. I would like to consume lower energy at home: & 0 (0\%) - 1 (2.24\%) - 2 (13.43\%) - 3 (51.49\%) - 4 (32.84\%) & P.2 \\
          \addlinespace
          &3. I would like to consume energy at home more efficiently: & 0 (0\%) - 1 (1.45\%) - 2 (13.04\%) - 3 (47.83\%) - 4 (37.68\%) & P.3 \\
          \addlinespace
          &4. I would like to make more efficient usage for the following reason: & Reduce my energy bill (66.66\%) - Contribute to the grid reliability, e.g. prevent a blackout (17.64\%) - Protect the environment (82.35\%) - Others do, so I do (9.80\%) - Others do not, so I do (1.96\%) & P.4 \\
          \addlinespace
          &5. I would like to use the following means to make a more efficient residential usage of energy: & Lowering my consumption of appliances (66.66\%) - Shifting the consumption of appliances at different times (e.g. during off-peak night times) (62.74\%) & P.5 \\
          \addlinespace
          &6. Which of the following would you, as a result of the automated control of residential appliances for a more efficient energy usage, find create discomfort: & Feeling cold in cold winters or feeling warm in warm summers (56.86\%) - Extra cost for special equipment and appliances (31.37\%) - Changing my overall lifestyle at home (35.29\%) - Doing my daily residential activities at different and maybe undesirable times (45.09\%) & P.6 \\
          \addlinespace
          &7. I would like to accept discomfort to make more efficient energy usage: & 0 (9.8\%) - 1 (21.6\%) - 2 (27.5\%) - 3 (35.3\%) - 4 (5.8\%) & P7\label{q:lambdaQuestion} \\
          \addlinespace
          &8. I would like to sacrifice energy efficiency to experience a low discomfort: & 0 (6.52\%) - 1 (15.22\%) - 2 (43.48\%) - 3 (34.78\%) - 4 (0\%) & P.8 \\
          \addlinespace
          &9. I would like to be more energy efficient if I know that others are more energy efficient as well: & 0 (8.70\%) - 1 (21.74\%) - 2 (28.26\%) - 3 (36.96\%) - 4 (4.35\%) & P.9 \\
          \addlinespace
          &10. I can accept a discomfort caused by energy efficiency if others can accept it as well: & 0 (10.87\%) - 1 (21.74\%) - 2 (30.43\%) - 3 (30.43\%) - 4 (6.52\%) & P.10 \\
          \addlinespace
          &11. I would not like to be energy efficient if others are not energy efficient as well: & 0 (34.78\%) - 1 (34.78\%) - 2 (26.09\%) - 3 (4.35\%) - 4 (0\%) & P.11 \\
          \addlinespace
          &12. I would not like to experience higher discomfort by energy efficiency if others do not experience higher as well: & 0 (26.09\%) - 1 (41.30\%) - 2 (26.09\%) - 3 (6.52\%) - 4 (0\%) & P.12 \\
          \addlinespace
          &13. I would like to allow technology to schedule a more efficient energy usage of my appliances: & 0 (2.17\%) - 1 (23.91\%) - 2 (54.35\%) - 3 (19.57\%) - 4 (0\%) & P.13 \\
          \addlinespace
          &14. I am willing to scheduler the use of appliances to make more efficient energy usage: &  0 (0\%) - 1 (6.52\%) - 2 (23.91\%) - 3 (56.52\%) - 4 (13.04\%) & P.14 \\
          \addlinespace
         &15. Scheduling of appliance to make more efficient energy usage best works for me: & 30? ahead (25.49\%) - 1 hour ahead (23.52\%) - 3 hours ahead (29.41\%) - 6 hours ahead (15.68\%) - 12 hours ahead (19.60\%) - 24 hours ahead (9.80\%) - 1 week ahead (7.8\%) & P.15 \\
          \addlinespace
          &16. For which discomfort level would you like to overtake control back over an appliance schedule for an efficient energy usage: &   0 (0\%) - 1 (26.09\%) - 2 (39.13\%) - 3 (32.61\%) - 4 (2.17\%) & P.16 \\
          \addlinespace
          \cline{1-4}
          \bottomrule
     \end{tabularx}
%%%%
%%%%
%%%%
%\twocolum
%%%%
%%%%
%%%%
\end{document}